\documentclass[journal]{IEEEtran}
\usepackage{cite}

\ifCLASSINFOpdf
\else
\fi
\usepackage{amsmath}
\usepackage{amssymb}

\usepackage{graphicx}

\graphicspath{{./Figures/}}
\DeclareGraphicsExtensions{.jpg}

\usepackage{url}

\usepackage[utf8]{inputenc}
\usepackage{subfiles}
\usepackage{xspace}

\usepackage[ruled,linesnumbered]{algorithm2e}
\usepackage{setspace}
\usepackage{enumerate}
\usepackage{amssymb, amsmath}
\usepackage{here}
\usepackage{microtype}

\usepackage{epstopdf}
\usepackage{bm}

\usepackage{mathrsfs}

\usepackage{tagging} 

\usepackage{breqn} 

\usepackage{pgf}
\usepackage{tikz}
\usepackage{tikz-network}
\usetikzlibrary{arrows,automata}
\usetikzlibrary{positioning}

\tikzset{
    state/.style={
           rectangle,
           rounded corners,
           draw=black, thick,
					 node distance=1.5cm,
           minimum height=2.5em,
					 minimum width=2.5em,
           inner sep=2pt,
           text centered,
           },
	  dots/.style={
					 node distance=1.5cm,
           minimum height=2em,
           inner sep=2pt,
           text centered,
           }
}

\usepackage{amssymb} 
\usepackage{xcolor,colortbl}
\usepackage{privvizNew}

\usepackage{paralist}

\usepackage{verbatim} 

\usepackage[n, operators, sets, landau, logic, ff, mm, probability]{cryptocode}

\usepackage{amsthm}
\newtheorem{definition}{Definition}
\newtheorem{theorem}{Theorem}

\newcommand{\Unlink}{\Gamma^{Unlink}_{\A, G'}(\lambda)}

\newcommand{\DC}{\mathcal{DC}}
\newcommand{\SM}{\mathcal{SM}}
\newcommand{\Ack}{\mathrm{Ack}}
\newcommand{\Ackring}{\mathrm{Ack_S}}

\newcommand{\EOR}{t}

\newcommand{\Lact}{\mathrm{L_{act}}}
\newcommand{\Lcand}{\mathrm{L_{rem}}}
\newcommand{\PRF}[1]{\mathrm{PRF}_{k_{#1}}(t)}
\newcommand{\nmin}{\mathrm{N}_{min}}

\newcommand{\ID}{\mathrm{ID}}
\newcommand{\cat}{\mathrm{\vert \vert}}

\newcommand{\A}{\mathcal{A}}

\newcommand{\papertitle}{AggFT\xspace}

\begin{document}

%
\title{\papertitle: Low-Cost Fault-Tolerant Smart Meter Aggregation with Proven Termination and Privacy}
%
%
%

\author{Günther~Eibl, \thanks{G. Eibl is with the Center for Secure Energy Informatics, Salzburg University of Applied Sciences, Puch bei Hallein, Austria.}
        Sanaz~Taheri-Boshrooyeh,~\IEEEmembership{Member,~IEEE,} \thanks{S. Taheri partly did this work at Koç University, İstanbul, Turkey.}%
        Alptekin~Küpçü,~\IEEEmembership{Senior Member,~IEEE} \thanks{A. Küpçü is with the Cryptography, Security and Privacy Research Group, Koç University, İstanbul, Turkey.}%
    }
\maketitle

\begin{abstract}
Smart meter data aggregation protocols have been developed to address rising privacy threats against customers' consumption data. However, these protocols do not work satisfactorily in the presence of failures of smart meters or network communication links.
In this paper, we propose a lightweight and fault-tolerant aggregation algorithm that can serve as a solid foundation for further research. We revisit an existing error-resilient privacy-preserving aggregation protocol based on masking and improve it by: (i) performing changes in the cryptographic parts that lead to a reduction of computational costs, (ii) simplifying the behaviour of the protocol in the presence of faults, and showing a proof of proper termination under a well-defined failure model, (iii) decoupling the computation part from the data flow so that the algorithm can also be used with homomorphic encryption as a basis for privacy-preservation. To best of our knowledge, this is the first algorithm that is formulated for both, masking and homomorphic encryption. (iv) Finally, we provide a formal proof of the privacy guarantee under failure. The systematic treatment with strict proofs and the established connection to graph theory may also serve as a starting point for possible generalizations and improvements with respect to increased resilience. 


\end{abstract}

\usetag{long} 


%
\IEEEpeerreviewmaketitle

\begin{taggedblock}{short}
\vspace{-0.7cm}
\end{taggedblock}

\section{Introduction} \label{sec:introduction}
Future energy systems will turn into increasingly complex systems with a demand for improved prediction and monitoring of produced and consumed aggregated energy. Smart meters may play a vital role by providing
data aggregated over user groups for regular time intervals needed for use cases such as forecasting or settlement. This paper focuses on provision of such aggregated data in a privacy preserving way, i.e. without revealing data from a single household. The billing use case, where data of a single household is needed, 
with 
pricing models like time of use-, block-, peak-based or dynamic tariffs \cite{Azarova19a,Bloch19a} is not considered here, since aggregation may only be appropriate for simple, consumption-based pricing models. For methods tackling such pricing models we refer to \cite{Rial11a, Rial16a}. 

A large body of research is dedicated to privacy-preserving methods of smart meters data aggregation \cite{Shi11a, Kursawe11a, Engel17a, Danezis13a}. However, one main real-world factor that is mostly overlooked in those studies is the failure of network connections or meters. A failing node or link can block the calculation of the correct aggregate. The recovery solutions are communication-costly or computationally-heavy where they require one extra round of execution or rely on a trusted third party. 

The lack of research in fault-tolerant aggregation protocols has already been recognized in \cite{Knirsch16a} (see also Section \ref{sec:LitSurvey}). Consequently, a cheap and error-resilient aggregation algorithm based on masking was developed in \cite{Knirsch16a}. However, the proposed solution shows deficiencies in capturing all the network failures, leading to cases where the correct calculation of the aggregate cannot be guaranteed. 

Realizing that the solution of \cite{Knirsch16a} has deficiencies, we greatly extend it by improving and generalizing the data aggregation protocol. 
In this paper, we propose \papertitle, which is a protocol for aggregation over a group of households  that is  efficient for the computationally constrained smart meters, and is fault-tolerant with provision of well-founded proofs for termination and privacy. We make a step towards a real-world deployment by naturally dealing with failures of network connections and smart meters.
Our new protocol has the following contributions:  
\begin{enumerate}
    \item \textbf{Light Computation:} We create a solution with low computational cost due to masking.
    \item \textbf{Fault-Tolerance and Guaranteed Termination:} The new algorithm handles all communication errors and failing parties, without needing an additional execution round or a trusted third party. Termination is guaranteed and formally proven under clear and reasonable assumptions about failures. 
    \item 
    \textbf{Modular Architecture and Solution Generality: } 
    Our proposed aggregation protocol follows a modular architecture in which the data-flow is decoupled from data processing and computation, and each is handled in a different layer. 
    Our computation-agnostic data-flow features guaranteed termination regardless of the underlying privacy-preserving aggregation mechanism. We illustrate this capability by using both masking and homomorphic encryption as two different privacy-preserving base methods. To the best of our knowledge, this protocol is the first smart meter aggregation protocol that can be used with either masking or homomorphic encryption. 
    \item \textbf{Formal Privacy Definition and Proof:} We propose a formal definition signifying measurement privacy against an adversarially-enforced network failure. Our proposed definition is general and can be employed by other aggregation protocols concerning user privacy in a faulty network.
    Furthermore, 
    we prove privacy formally with a game-based cryptographic proof, which also clarifies the privacy-impact of failed links.
\end{enumerate}

\section{Related Work} \label{sec:preliminaries}
\subsection{Existing Aggregation Algorithms} \label{sec:LitSurvey}
\begin{taggedblock}{long}
Aggregation protocols aim at privately  summing the values of a given set of smart meters. The main methods (sometimes combined) are: masking \cite{ Kursawe11a, Erkin12b, Acs11a, Marmol13a, Knirsch16a, Bonawitz17a}, additive homomorphic encryption \cite{Shi11a, Engel17a, Li10a, Chan12a, Rastogi10a, Erkin12b},
and  multiparty computation with secure secret sharing \cite{Danezis13a, Bonawitz17a, Mustafa19a}. 
\begin{taggedblock}{long}
Differential privacy is typically applied on top of these aggregation methods  \cite{Acs11a, Shi11a, Rastogi10a}, thereby adding a privacy guarantee to the aggregated value.
\end{taggedblock}

Many papers propose privacy-preserving protocols  without providing a mechanism that works in case of faulty smart meters or non-working communication links \cite{Shi11a, Kursawe11a, Engel17a, Danezis13a}. Only a small number of papers discuss fault-tolerance (typically only failures of smart meters but not communication links).
In \cite{Li10a}, aggregation using homomorphic encryption 
proceeds along an aggregation tree. A failing node would result in loss of values for all meters of the subtree rooted at that node.
In \cite{Chan12a} this problem is solved and the solution can  also be applied to the methods of \cite{Shi11a} and \cite{Rastogi10a}. 
\begin{taggedblock}{long}
The solution of \cite{Chan12a} is intended to work for quite different protocols that (i) use homomorphic encryption method enhanced with differential privacy and (ii) assume a star topology where the nodes only communicate with the aggregator. The key to achieving fault-tolerance is to organize the nodes into a tree of user groups. Each user is contained in several blocks. When a user fails, aggregation must be done over disjoint blocks that cover the functioning users.
\end{taggedblock}
However, to cover all functioning nodes, each node would need to perform roughly $\lfloor \log_2(n) \rfloor +1$ costly homomorphic encryption operations. This computational overhead for devices with low computational power, such as smart meters, should be avoided. 
\begin{taggedblock}{long}
In \cite{Rastogi10a}, fault-tolerance is achieved by employing a \textit{thresholded} Paillier-cryptosystem which works if enough parties remain. This solution can of course not be applied to the cheaper protocols that rely on masking.
\end{taggedblock}
The protocol of \cite{Erkin12b} combines homomorphic encryption with masking in the exponent of Paillier's encryption scheme \cite{Paillier99a}. In case of malfunctioning smart meters, missing random numbers are proposed to be retrieved by additionally contacting a trusted third party (the manufacturer) that can provide the missing information in a privacy-preserving way.

In \cite{Acs11a}, masking is employed with pairwise random shares that cancel each other in clusters of smart meters. Since missing shares would destroy the result, the corresponding counterparts need to be queried in an additional round.
\begin{taggedblock}{long}
The paper nicely extends the basic masking procedure and achieves a differentially private aggregate by adding noise in a distributed way. With this second round it not only ensures that masking still works but also ensures that differential privacy is maintained.
\end{taggedblock}
In \cite{Marmol13a} all smart meters send their masked values directly to the energy supplier (ES), which knows the sum of the shares.
If a smart meter fails, the initial exchange of the random shares must be triggered again. Lost messages are treated using acknowledgement (ACK) signals similar to \cite{Knirsch16a}, and require not only additional steps but also that the key aggregator and the ES do not collude.
\begin{taggedblock}{long}
Masking is also the basic method applied in \cite{Danezis13a}: there the random values are hashes that are secretly shared with additional parties, called authorites. This method also allows for various other computations than just the sum and privacy is provided if at least one authority is not corrupted.
\end{taggedblock}

In the context of federated learning with sensitive data \cite{Bonawitz17a}, masking is combined with secret sharing to handle dropping users. In this case, a recovery round is needed, and a second masking procedure is applied to preserve privacy of the dropping users in the recovery phase.
Secure multiparty computation using threshold Shamir secret sharing with fault-tolerance is proposed in \cite{Mustafa19a}. While computations for smart meters are cheap, most computations are done at DCCs (Data Communication Companies). 
Trust in the DCCs is needed, since  
privacy is not guaranteed if all DCCs collude.

In all these papers, fault tolerance needs additional effort or trust in additional parties. Moreover, informal privacy discussions are still more common \cite{Li10a, Knirsch16a, Erkin12a, Engel17a, Acs11a, Kursawe11a} than formal proofs. Only in \cite{Knirsch16a}, aggregation can be done in presence of failures without any of these requirements. However, no proof for termination exists and indeed some failures lead to an incomplete round (see Section \ref{sec:improveFailureResistance}). 

\end{taggedblock}
\begin{taggedblock}{short,medium}
Aggregation protocols aim at privately  summing the values of a given 
set of smart meters. Methods are based on masking \cite{ Kursawe11a, Erkin12b, Acs11a, Marmol13a, Knirsch16a, Bonawitz17a}, additive homomorphic encryption \cite{Shi11a, Engel17a, Li10a, Chan12a, Rastogi10a, Erkin12b} or  multiparty computation with secure secret sharing \cite{Danezis13a, Bonawitz17a, Mustafa19a} and often combined. 
Many papers propose privacy-preserving protocols  without providing a mechanism that works in case of faulty smart meters or non-working communication links \cite{Shi11a, Kursawe11a, Engel17a, Danezis13a, Li10a}. 

Only a small number of papers discuss fault-tolerance (typically only failures of smart meters but not communication links).
In \cite{Chan12a} this problem is solved and the solution can  also be applied to the methods of \cite{Shi11a} and \cite{Rastogi10a}. 
However, covering all functioning nodes requires each computationally restricted node to perform roughly $\lfloor \log_2(n) \rfloor +1$ costly homomorphic encryption operations.
Homomorphic encryption with can be combined with masking in the exponent of Paillier's  encryption scheme \cite{Erkin12b}. 
In case of malfunctioning smart meters, missing random numbers are proposed to be retrieved by additionally contacting a trusted third party (the manufacturer) that can provide the missing information in a privacy-preserving way.

In \cite{Acs11a}, masking is employed with pairwise random shares that cancel each other in clusters of smart meters. Since missing shares would destroy the result, the corresponding counterparts need to be queried in an additional round.
In \cite{Marmol13a} all smart meters send their masked values directly to the energy supplier (ES), which knows the sum of the shares.
If a smart meter fails, the initial exchange of the random shares must be triggered again. Lost messages are treated using acknowledgement (ACK) signals similar to \cite{Knirsch16a}, and require not only additional steps but also that the key aggregator and the ES do not collude.

In the context of federated learning with sensitive data \cite{Bonawitz17a}, masking is combined with secret sharing to handle dropping users. In this case, a recovery round is needed, and a second masking procedure is applied to preserve privacy of the dropping users in the recovery phase.
Secure multiparty computation using threshold Shamir secret sharing with fault-tolerance is proposed in \cite{Mustafa19a}. While computations for smart meters are cheap, most computations are done at DCCs (Data Communication Companies).
Trust in the DCCs is needed, since  
privacy is not guaranteed if all DCCs collude.

In all these papers, fault tolerance needs additional effort \cite{Chan12a, Acs11a,Marmol13a,Bonawitz17a} or trust in additional parties \cite{Erkin12b,Mustafa19a}. 
Moreover, informal privacy discussions are still more common \cite{Li10a, Knirsch16a, Erkin12a, Engel17a, Acs11a, Kursawe11a} than formal proofs. 
Only in \cite{Knirsch16a}, aggregation  can be done in presence of failures without any of these requirements. 
However, no proof for termination exists and some failures lead to an incomplete round (see Section \ref{sec:improveFailureResistance}). 
\end{taggedblock}

\subsection{Error-Resilient Masking Algorithm from \cite{Knirsch16a}} \label{sec:existingAlgo}

The intended flow of the aggregation protocol developed in \cite{Knirsch16a} is shown in Figure \ref{fig: newprot}, the algorithm itself is described in Algorithm  \ref{fig: algoExisting}, omitting the temporal aggregation part.
\begin{taggedblock}{long}
In each round each smart meter $\SM_i$ masks its measurement $m_{i,t}$ by adding a random share $s_{i,t}\in \mathbb{Z}_k$ and a share $s_i^0$ that is only known by $\SM_i$ and the $\DC$ (line 15). Since $s_{i,t}$ is only known to $\SM_i$, it hides (masks) the measurement $m_{i,t}$. If the $\DC$ calculates the sum of the masked measurements, it can subtract $\sum_i s_i^0$, since he has this information. However the sum of the shares $s_{i,t}$ still masks the aggregate. This sum is calculated along the ring of smart meters without the $DC$ which only gets the sum at the end of a round.  
\end{taggedblock}

The intended flow proceeds along an ordered sending list $L=(\DC,1,\ldots,N,\DC)$ starting at the Data Concentrator ($\DC$). 
The $\DC$ sends its random share $S_0:=s_{0,t} \in \mathbb{Z}_k$ to smart meter 1 ($\SM_1$) (line 8). $\SM_1$ gets active, masks its measurement $m_{1,t}$ (line 15) using its share $s_{1,t}$ and adds its share to the one obtained from the $\DC$ (line 16). The static shares $s_i^0$ are shared with the $DC$ and intended as a means against eavesdroppers.
Then $\SM_1$ sends the masked measurement $\tilde{m}_{1,t}$ to the $\DC$ (line 17) and the updated sum of shares $S_1=S_0+s_{1,t}$ to the the next party in $L$ ($\SM_2$), which gets active.   
This procedure continues until the last smart meter $\SM_N$ sends the sum of the shares $S_N$ to the $DC$ (line 20). The $DC$ subtracts $S_N$ from the sum of the masked measurements which yields the desired aggregate (line 28).

\def\xDC{3}
\def\yDC{2}
\def\ySM{0}

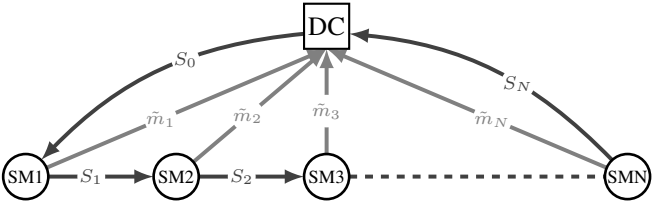
\begin{figure}
	\centering
	\begin{tikzpicture}


\Text[x=\xDC,y=\yDC] {DC};
\draw[black,thick] (\xDC-0.3,\yDC+0.3) rectangle (\xDC+0.3,\yDC-0.3);

\Vertex[x=-1, y=\ySM,color=white, label=SM1]{SM1}
\Vertex[x=1, y=\ySM, color=white, label=SM2]{SM2}
\Vertex[x=3, y=\ySM, color=white, label=SM3]{SM3}
\Vertex[x=7, y=\ySM, color=white, label=SMN]{SM5}

\coordinate (DC1) at (\xDC,\yDC-0.3);
\coordinate (DCleft) at (\xDC-0.3,\yDC-0.1);
\coordinate (DCright) at (\xDC+0.3,\yDC-0.1);

\Edge [color=gray, Direct, label=$\tilde{m}_1$,distance=.4](SM1)(DC1)
\Edge [color=gray, Direct, label=$\tilde{m}_2$,distance=.4](SM2)(DC1)
\Edge [color=gray, Direct, label=$\tilde{m}_3$,distance=.4](SM3)(DC1)
\Edge [color=gray, Direct, label=$\tilde{m}_N$,distance=.4](SM5)(DC1)

\Edge [Direct, label=$S_0$,distance=.4, bend=-20](DCleft)(SM1)
\Edge [Direct, label=$S_1$,distance=.4](SM1)(SM2)
\Edge [Direct, label=$S_2$,distance=.4](SM2)(SM3)
\Edge [style=dashed](SM3)(SM5)
\Edge [Direct, label=$S_N$,distance=.4, bend=-20](SM5)(DCright)

\end{tikzpicture}
	\caption{Illustration of the information flow of the masking protocol: it starts at the $\DC$, proceeds along the ring of smart meters and ends at the $\DC$. The sum of shares $S$ is computed along the ring. When a smart meter gets active it sends its masked value $\tilde{m}$ directly to the $\DC$ (gray arrows). In practice, the arrows point in both directions because each message is directly answered by a corresponding ACK-message.}
	\label{fig: newprot}
	\vspace*{-3ex}
\end{figure}


\begin{algorithm}[ht]
	\setstretch{0.8}
	\vspace{1mm}
	\textbf{Initialization}\\
	Provide sending list $L$ to all smart meters and to $\DC$\\
	$\forall i$: $\DC$ sends static share $s_i^0$ to $\SM_i$, $S^0=\sum_i s_i^0$\\
	\vspace{-2mm}
	\hrulefill \\
	\While{$t \leq T$}{
		\textbf{One round of reading }$t$\\
		All participating parties $i$ generate a random $s_{i,t}$\\
		DC sends $M_{0 \rightarrow 1} =(s_{0,t} \cat \ID_\DC)$ to $\SM_1$, $n=1$\\
		\While{$\mathrm{no} \ M^{\Ack}_{0\leftarrow n} \ \mathrm{from} \ \SM_n \ \mathrm{within} \ \Delta t$}{
			$n=n+1$, DC sends $M_{0\rightarrow n}$ to next $\SM_n$.
		}
		$i=n$, $p=0=\DC$, $n=i+1$\\
		\While{$i \leq N$}{
			\tcc{Only $\SM_i$ acts}
			Send an Ack-signal $M^{\Ack}_{p \leftarrow i}$ to $L(p)$\\
			$\tilde m_{i,t}= m_{i,t}+s_{i,t}+s_i^0 \mod k$\\
			$S_i= S_{p}+s_{i,t} \mod k$\\
			Send 
			$(\tilde m_{i,t} \cat H(m_{i,t}) \cat H(s_{i,t}) \cat t)$ to $\DC$\\
			
			Send 
			$S_i$ to $L(n)$.\\
			\While{$\SM_i \ \mathrm{does \ not \ get \ } M^{\Ack}_{n\to i} \ \mathrm{within} \ \Delta t$}{
				$n=n+1$, send $S_i$ to $L(n)$
			}
			$p=i$, $i=n$
		}
		\vspace{-2mm}
		\hrulefill \\
		\textbf{Aggregation}\\
		\tcc{Only $\DC$ acts}
		Check if  $H(S_N) = \prod_i H(s_{i,t})$\\
		$\forall \tilde m_{i,t}$: check $H(\tilde m_{i,t}) = H(m_{i,t}) H(s_{i,t}) H(s_i^0)$\\
		Calculate aggregate $A_t = \sum_i \tilde m_{i,t} - S_N-S^0$\\
		Increase $t$
	}
    \caption{Aggregation algorithm of \cite{Knirsch16a}}  \label{fig: algoExisting}
\end{algorithm}

The key to privacy is the fact that (i) the next ($n$) smart meters in the sending list $L$ only gets random shares (line 18) and (ii) the $\DC$ only gets masked measurements $\tilde{m}_{i,t}$ and hashes (line 17). The $\DC$ can only reveal the aggregate by subtracting $S_N$. The key to fault-tolerance is that each smart meter tries to contact the next smart meter in the list (lines 18-22). If the link and the next smart meter work, the successfully contacted smart meter sends an ACK signal back (line 14) such that the currently active smart meter checks that the next one takes over (line 19) and that it can get inactive. 

In masking protocols, a single inconsistency between the shares added along the ring and the shares added to the measurements completely destroys the result. In order to check consistency, hashes of the measurements $H(m_{i,t})$ and the shares $H(s_{i,t})$ are introduced. A missing link of the active smart meter $\SM_i$ to the $\DC$ and consistency can then be detected by the check of the hashes of the shares in lines 26 and 27.
However, this failure is only detected but not corrected.





\section{\papertitle} \label{sec:protocol}
While Algorithm \ref{fig: algoExisting} makes an important step towards fault-tolerance, a re-investigation shows several issues; problems were partly arising due to missing assumptions about the errors and missing formal proofs. 

\subsection{Assumptions}

To protect privacy against external attackers, our solution needs secure and authenticated channels that connect the smart meters to the $\DC$. Since during  smart grid deployment, the $\DC$ initially configures the smart meters (e.g., setting up reporting procedure and frequency), secure and authenticated channels can easily be established using only symmetric-key authenticated encryption with pre-shared keys between each $\SM_i$ and $\DC$ without requiring heavy public-key cryptography.

The common assumption about failures here will be that a smart meter or  a communication link is either working correctly (on) or not (off) during a single aggregation round. 
Link failures are assumed bi-directional; therefore, the link-structure during a round can be described as an undirected graph $G=(V, E)$, as shown in Figure \ref{fig:exNetwork0}. The vertices $V$ correspond to the entities involved in the aggregation protocol, lines describe the edges (links) $E$. Solid lines are the links $E'\subseteq E$ working in round $t$, dashed lines are the links that fail. The check that a link is on can be done by sending a message, which is answered by an ACK return message. If a link is on, both the original and the return messages get through. Therefore both the receiver and the sender know, whether the link is on or not. If a link or a smart meter is off, 
the sender 
does not get an ACK within a time period $\Delta T$, and therefore also knows that the link or the receiver is off. The improved Algorithm \ref{fig: algoGeneralV02} does not specify $\Delta T$ and assumes that a proper $\Delta T$ can be empirically set. The $\DC$ is assumed to be always on.

\def\xDC{4}
\def\yscale{0.8}
\def\bend{50}
\def\bendz{15}


\subsection{Failure Resistance and Simplifying the Flow}
\label{sec:improveFailureResistance}

\papertitle, as illustrated in Algorithm \ref{fig: algoGeneralV02}, is designed to proceed from the $\DC$ along the ring of smart meters and back to the $\DC$. Similarly to Algorithm \ref{fig: algoExisting}, during proceeding along the ring, when a party is \textit{active}, it performs some computations and then activates the next party in the list.


\begin{algorithm}[htb]
  \SetAlgoLined\DontPrintSemicolon
  \SetKwFunction{algo}{main}
  \SetKwFunction{proc}{sendFinalMessage}
  \SetKwFunction{proce}{checkLast}
  \SetKwProg{myalg}{Algorithm}{}{}
  \myalg{\algo{}}{
   	    \tcc{Sending of initial data:}
		\color{blue}All $\SM_i$ send $\langle t,i,\mathrm{data}_i\rangle$ to $\DC$ and get inactive\color{black}\;
		$\DC$ sets $\Lcand=( i\vert \ \mathrm{data}_i$ was received$)$\;
		\textbf{If} ($\vert \Lcand \vert \geq \nmin$): stop round $t$.\; 
		$\DC$ sets $\Lact=\{\}$\;
		\color{blue}$\DC$ calculates $S$\color{black}\;
		\tcc{Start of aggregation}
		$\DC$ chooses first entry $i$ from $\Lcand$\;
		$\DC$ sends $\langle S, \ \Lcand, \ \Lact\rangle$ to  $\SM_i$\;
		\tcc{$\SM_i$ gets active \& takes over}
		$\SM_i$ sends $\Ackring$ back\;
		\color{blue}$\SM_i$ calculates $S_{i}$ sets $S=S_i$\color{black}\;
		$\SM_i$ sets $\Lact=\Lact \cup \{i\}, \ \Lcand=\Lcand \setminus \{i\}$\;
		$isLast = $ \proce{$\Lcand, \Lact, \nmin$}\;
		$tryNext=\neg isLast$\;
		\While{tryNext=true}
		{
    	    $\SM_i$ chooses next $j\in \Lcand$\;
    	    $\SM_i$ sends $\langle S, \ \Lcand, \ \Lact\rangle$ to  $\SM_j$\;  
    	    \eIf{$\Ackring$ \rm{from} $\SM_j$}
        	    {
        	    $tryNext=false$\;
        	    }
        	    {
        	        $\SM_i$ sets $\Lcand=\Lcand \setminus \{j\}$\;
        	        $isLast=$ \proce{$\Lcand, \Lact, \nmin$}\;
                    $tryNext=\neg isLast$\;


        		}
		}
    		\If{$isLast=true$}
    		{
    		\proc{$S, \Lcand, \Lact$}
    		}
    		\tcc{$\SM_i$ gets inactive}

		\tcc{Finish the round at $\DC$:}
		\If{$\DC \ \mathrm{receives} \ \langle \EOR, \ S, \ \Lact  \rangle$}
		{
		\color{blue}$\DC$ \rm{calculates the aggregate using} $S$ and $\Lact$\color{black}\;
	    }
  \KwRet\;}
  
  \setcounter{AlgoLine}{0}
  \SetKwProg{myproc}{Procedure}{}{}
  \myproc{\proc{$S, \Lcand, \Lact$}}{
  \If{$\vert \Lcand \cup \Lact \vert < \nmin$}
  {
  $S=\{\}, \ \Lact=\{\}$
  }
  send $\langle \EOR, \ S, \ \Lact  \rangle$ to $\DC$\;
  \KwRet\;}

  \setcounter{AlgoLine}{0}
  \SetKwProg{myproc}{Procedure}{}{}
  \myproc{\proce{$\Lcand, \Lact, \nmin$}}{
        $isLast=(\Lcand = \{ \})  \lor   (\vert \Lcand \cup \Lact \vert < \nmin)$\;
  \KwRet\;}

  \caption{\papertitle} \label{fig: algoGeneralV02}
 \end{algorithm}

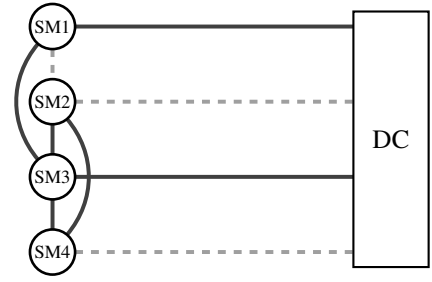
\begin{figure}[htb]
	\centering
	\begin{tikzpicture}


\Text[x=4.5,y=-1.5] {DC};
\draw[black,thick] (\xDC,0.2) rectangle (\xDC+1,-3.2);
\Vertex[color=white, y=0, label=SM1]{SM1}
\Vertex[y=-1, color=white, label=SM2]{SM2}
\Vertex[y=-2,color=white, label=SM3]{SM3}
\Vertex[y=-3, color=white, opacity=0.7, label=SM4]{SM4}

\coordinate (DC1) at (\xDC,0);
\coordinate (DC2) at (\xDC,-1);
\coordinate (DC3) at (\xDC,-2);
\coordinate (DC4) at (\xDC,-3);

\Edge (SM1)(DC1)
\Edge [opacity=.5, style=dashed](SM2)(DC2)
\Edge (SM3)(DC3)
\Edge [opacity=.5, style=dashed](SM4)(DC4)

\Edge [opacity=.5, style=dashed](SM1)(SM2)
\Edge [bend=-40](SM1)(SM3)
\Edge (SM2)(SM3)
\Edge [bend=40](SM2)(SM4)
\Edge (SM3)(SM4)

\end{tikzpicture}
\caption{Example network consisting of 4 smart meters and the data concentrator used to demonstrate arising problems. Connections that are on and off at round $t$ are solid and dashed, respectively.}
\label{fig:exNetwork0}
\vspace*{-2ex}
\end{figure}

The most important feature of \papertitle is the following: the smart meters do not send their data (masked measurements) in the loop but right at the beginning (line 2). By only considering smart meters where the message gets through (line 3), this subtle change removes the uncertainty about (i) all the links to the $\DC$ and (ii) failures of smart meters. Since the smart meters and the link to the $\DC$ work at the beginning, the assumption about link-failures formally ensures that this also holds for the rest of the round. For the example network in Figure \ref{fig:exNetwork0}, it is therefore known from the beginning that smart meters $\SM_2$ and $\SM_4$ can not send their masked measurement to the $\DC$ in this round. Since they can therefore not \textit{contribute} to the aggregate, it makes sense to remove smart meters $\SM_2$ and $\SM_4$ at the beginning, so they can never get active. 

Sending the data to the $\DC$ right at the beginning also has an important consequence for the termination of \papertitle. For the example network in Figure \ref{fig:exNetwork0}, Algorithm \ref{fig: algoExisting} would proceed from the $\DC$ via $\SM_1$ and $\SM_3$ to $\SM_4$. However, it does not terminate, because $\SM_4$ has no instruction of how to send the sum of the shares $S_4$ to the $\DC$ when this link fails. In contrast, \papertitle excludes $\SM_2$ and $\SM_4$ right at the beginning so that this case can not happen. While this is only an intuitive example, the termination of \papertitle will be proven in Section \ref{sec:termination}.  
Note that the proof of termination heavily influenced the resulting algorithm since the idea for this change came from the desire to rule out this situation during working on the proof. This demonstrates the necessity and benefit of formally proving desired properties of the protocol.

\papertitle also ensures that for privacy reasons at least $\nmin$ smart meters contribute to the aggregate. Since later it can be proven that all active smart meters will contribute these are stored in $\Lact$. Only acting in the initialization part (especially line 2) is not considered as active. Note that $\Lact$ needs to be updated in each step, because the currently active smart meter does not know, if the next smart meter can be reached (there is still uncertainty about the links between smart meters). Therefore, each active smart meter adds itself to $\Lact$ and removes itself from the list of remaining smart meters $\Lcand$ (line 11). The end of the round is then determined in procedure \texttt{checkLast} by checking that no smart meters remain to be contacted ($\Lcand=\{\}$). The $\nmin$-condition is checked in procedure \texttt{sendFinalMessage}, before the last smart meter sends its message to the $\DC$. Only if the condition is satisfied, the sum of shares $S$ is sent (otherwise $S$ is set as the empty set). The $\nmin$-checks in procedure \texttt{checkLast} only enables an earlier stop of the round, if it can be foreseen that the $\nmin$-check can not be satisfied any more.

\begin{taggedblock}{long}
We demonstrate, how Algorithm \ref{fig: algoGeneralV02} works in more detail 
by explanation for the example network in Figure \ref{fig:exNetworkExtended}. The resulting messages are illustrated in Figure \ref{fig:exampleSimplerAlgo}.
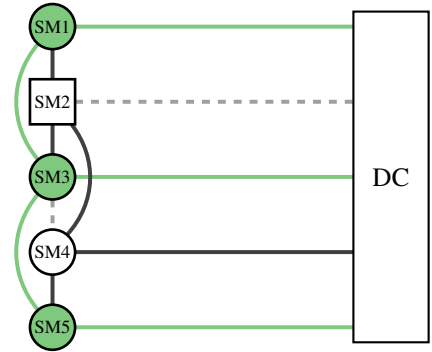
\begin{figure}[htb]
	\centering
	\begin{tikzpicture}


\Text[x=4.5,y=-2] {DC};
\draw[black,thick] (\xDC,0.2) rectangle (\xDC+1,-4.2);
\Vertex[RGB,color={127,201,127}, y=0, label=SM1]{SM1}
\Vertex[y=-1, color=white, label=SM2, shape = rectangle]{SM2}
\Vertex[y=-2, RGB,color={127,201,127}, label=SM3]{SM3}
\Vertex[y=-3, color=white, opacity=0.7, label=SM4]{SM4}
\Vertex[RGB,color={127,201,127}, y=-4, label=SM5]{SM5}

\coordinate (DC1) at (\xDC,0);
\coordinate (DC2) at (\xDC,-1);
\coordinate (DC3) at (\xDC,-2);
\coordinate (DC4) at (\xDC,-3);
\coordinate (DC5) at (\xDC,-4);

\Edge [RGB,color={127,201,127}](SM1)(DC1)
\Edge [opacity=.5, style=dashed](SM2)(DC2)
\Edge [RGB,color={127,201,127}](SM3)(DC3)
\Edge (SM4)(DC4)
\Edge [RGB,color={127,201,127}](SM5)(DC5)

\Edge (SM1)(SM2)
\Edge [bend=-40,RGB,color={127,201,127}](SM1)(SM3)
\Edge (SM2)(SM3)
\Edge [bend=40](SM2)(SM4)
\Edge [opacity=.5, style=dashed](SM3)(SM4)
\Edge [bend=-40,RGB,color={127,201,127}](SM3)(SM5)
\Edge (SM4)(SM5)

\end{tikzpicture}
\caption{Example network consisting of 5 smart meters and the $\DC$ used to demonstrate assumptions, algorithms and arising problems. Connections that are on and off at round $t$ are solid and dashed, respectively. Smart meters that will get active are filled, used links are green, unused links are black.}
\label{fig:exNetworkExtended}
\end{figure}

\begin{figure}
	\centering
	\begin{tikzpicture}

\node (DC) at (\xDC+0.5,-3*\yscale)   {DC};
\draw[black,thick] (\xDC,-0.5*\yscale) rectangle (\xDC+1,-7.5*\yscale);
\Vertex[RGB,color={127,201,127}, y=-1*\yscale, label=SM1]{SM1}
\Vertex[RGB,color={127,201,127}, y=-3*\yscale, label=SM3]{SM3}
\Vertex[color=white, y=-5*\yscale, label=SM4]{SM4}
\Vertex[RGB,color={127,201,127}, y=-7*\yscale, label=SM5]{SM5}

\coordinate (DC1) at (\xDC,-1*\yscale);
\coordinate (DC3) at (\xDC,-3*\yscale);
\coordinate (DC4) at (\xDC,-5*\yscale);
\coordinate (DC5) at (\xDC,-7*\yscale);

\Edge[Direct, label=1.$\tilde{m}_1$,distance=.4, bend=0](SM1)(DC1)
\Edge[Direct, label=1.$\tilde{m}_3$,distance=.4, bend=0](SM3)(DC3)
\Edge[Direct, label=1.$\tilde{m}_4$,distance=.4, bend=0](SM4)(DC4)
\Edge[Direct, label=1.$\tilde{m}_5$,distance=.4, bend=0](SM5)(DC5)

\Edge[Direct, label=2.$S_0$\textit{,L},distance=.6, bend=-20, position=above](DC1)(SM1)

\Edge[Direct, label=3.$S_1$\textit{,L},distance=.4, bend=(-1)*\bend](SM1)(SM3)
\Edge[Direct,label=4.,style=dashed,distance=.3,bend=-30](SM3)(SM1)

\Edge[Direct, label=5.$S_3$\textit{,L},distance=.4](SM3)(SM4)
\Edge[Direct, label=6.$S_3$\textit{,L},distance=.4, bend=-40](SM3)(SM5)
\Edge[Direct,label=7.,distance=.3,bend=-40,style=dashed](SM5)(SM3)

\Edge[Direct,label=8.$\EOR S_5$\textit{,L},distance=.6,bend=-20, position=below](SM5)(DC5)

\end{tikzpicture}
\label{fig:exampleSimplerAlgo}
\caption{
Demonstration of the simplified information flow: $\SM_2$ is removed, since only the smart meters that successfully send the masked measurements to the $\DC$ in step 1 can act later on. Dashed arrows are ack-signals that notify a successful reception of a message.
}
\end{figure}
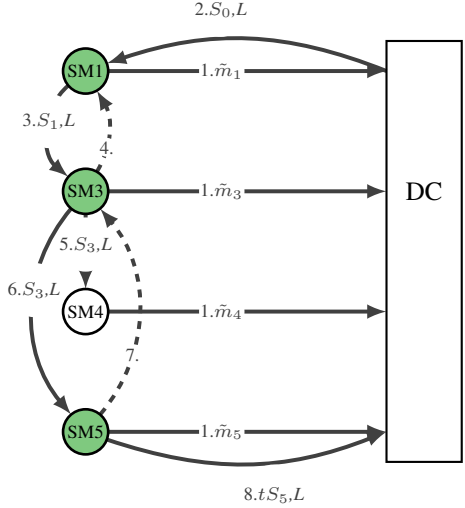

The immediate removal of smart meters without a connection to the $\DC$ leaves only the four smart meters in the list of remaining smart meters $\Lcand=\{1, 3, 4, 5\}$. When $\SM_1$ gets active, it sends an Ack-signal back (this special Ack-signal would not be necessary because it is sent to the $\DC$ and not a smart meter), does some computations, removes itself from $\Lcand$ (so $\Lcand=\{3, 4, 5\}$) and puts itself in the list of active smart meters $\Lact$ resulting in $\Lact=\{1\}$. Then it contacts the next member ($\SM_3$) of  $\Lcand$ which does the same  (so $\Lcand=\{4, 5\}$ and $\Lact=\{1,3\}$). However, since the link from the active smart meter $\SM_3$ to $\SM_4$ does not work, $\SM_4$ can not send an Ack back, $\SM_3$ therefore removes $\SM_4$ from $\Lcand$  (so $\Lcand=\{5\}$) and continues with $\SM_5$. After removing itself from $\Lcand$ and putting itself in the list of active smart meters $\Lact$ (so $\Lact=\{1,3,5\}$), smart meter $\SM_5$ realizes that it is the last smart meter, since $\Lcand$ is empty. Consequently it finishes the round by sending the needed information to the $\DC$ which can then calculate the aggregate. Since the number of contributing smart meters at the end of the round is at most $\vert \Lcand \cup \Lact \vert$, comparing this to $\nmin$ enables an earlier detection and finish of the round, if not enough smart meters will contribute to the aggregate. 

The simplification arising from sending the messages to the $\DC$ in line 2 can be illustrated by demonstrating the higher  complexity of an alternative algorithm that fixes the problems of Algorithm \ref{fig: algoExisting} in a more direct way. The resulting messages for a slightly different example are shown in Figure \ref{fig:complicated}.    
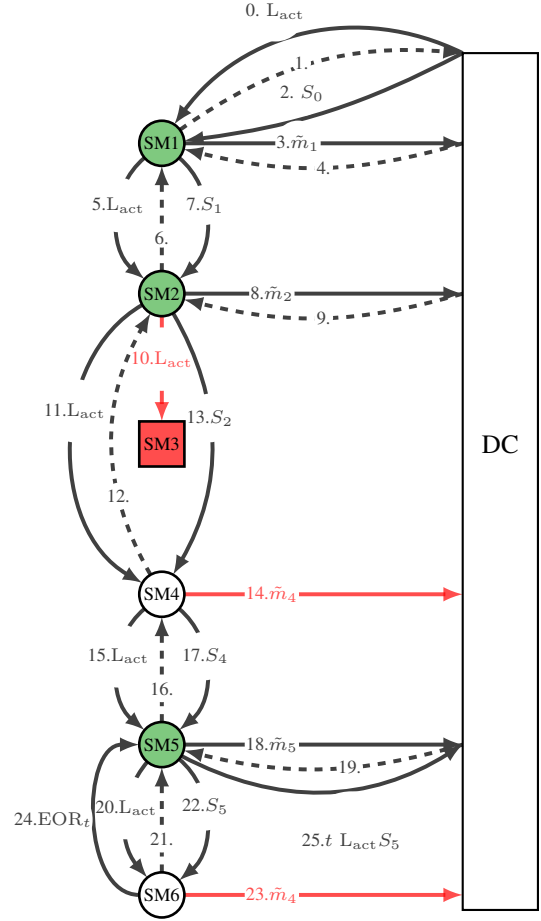
\begin{figure}
	\centering
	\begin{tikzpicture}

\node (DC) at (\xDC+0.5,-5)   {DC};
\draw[black,thick] (\xDC,0.2) rectangle (\xDC+1,-11.2);
\Vertex[RGB,color={127,201,127}, y=-1, label=SM1]{SM1}
\Vertex[RGB,color={127,201,127}, y=-3, label=SM2]{SM2}
\Vertex[y=-5, color=red, opacity=0.7, label=SM3, shape = rectangle]{SM3}
\Vertex[y=-7, color=white, label=SM4]{SM4}
\Vertex[RGB,color={127,201,127}, y=-9, label=SM5]{SM5}
\Vertex[y=-11, color=white, label=SM6]{SM6}

\coordinate (DC0) at (\xDC,0.2);
\coordinate (DC1) at (\xDC,-1);
\coordinate (DC2) at (\xDC,-3);
\coordinate (DC4) at (\xDC,-7);
\coordinate (DC5) at (\xDC,-9);
\coordinate (DC6) at (\xDC,-11);

\Edge[Direct, label=0. $\Lact$,distance=.6, bend=-40, position=above](DC0)(SM1)
\Edge[Direct, label=1.,distance=.5, bend=20, position=left, ,style=dashed](SM1)(DC0)
\Edge[Direct, label=2. $S_0$,distance=.6, bend=10, position=above](DC0)(SM1)

\Edge[Direct, label=3.$\tilde{m}_1$,distance=.4, bend=0](SM1)(DC1)
\Edge[Direct, label=4.,distance=.5, bend=\bendz,style=dashed](DC1)(SM1)
\Edge[Direct, label=5.$\Lact$,distance=.4, bend=(-1)*\bend](SM1)(SM2)
\Edge[Direct,label=6.,style=dashed,distance=.3,bend=0](SM2)(SM1)
\Edge[Direct,label=7.$S_1$,distance=.4,bend=\bend](SM1)(SM2)

\Edge[Direct, label=8.$\tilde{m}_2$,distance=.3, bend=0](SM2)(DC2)
\Edge[Direct, label=9.,distance=.5,bend=\bendz,style=dashed](DC2)(SM2)
\Edge[Direct, label=10.$\Lact$,distance=.4,bend=0, color=red, opacity=0.7](SM2)(SM3)
\Edge[Direct, label=11.$\Lact$,distance=.4, bend=-60](SM2)(SM4)
\Edge[Direct,label=12.,distance=.3,bend=30,style=dashed](SM4)(SM2)
\Edge[Direct,label=13.$S_2$,distance=.4,bend=30](SM2)(SM4)

\Edge[Direct, label=14.$\tilde{m}_4$,distance=.3, bend=0, color=red, opacity=0.7](SM4)(DC4)
\Edge[Direct, label=15.$\Lact$,distance=.4, bend=(-1)*\bend](SM4)(SM5)
\Edge[Direct,label=16.,distance=.3,bend=0,style=dashed](SM5)(SM4)
\Edge[Direct,label=17.$S_4$,distance=.4,bend=\bend](SM4)(SM5)

\Edge[Direct, label=18.$\tilde{m}_5$,distance=.3, bend=0](SM5)(DC5)
\Edge[Direct, label=19.,distance=.4,bend=\bendz,style=dashed](DC5)(SM5)
\Edge[Direct, label=20.$\Lact$,distance=.4, bend=-40](SM5)(SM6)
\Edge[Direct,label=21.,distance=.3,bend=0,style=dashed](SM6)(SM5)
\Edge[Direct,label=22.$S_5$,distance=.4,bend=\bend](SM5)(SM6)
\Edge[Direct,label=25.$\EOR \ \Lact S_5$,distance=.6,bend=-30, position=below](SM5)(DC5)

\Edge[Direct, label=23.$\tilde{m}_4$,distance=.3, bend=0, color=red, opacity=0.7](SM6)(DC6)
\Edge[Direct,label=24.$\mathrm{EOR}_t$,distance=.5,bend=90, position=left](SM6)(SM5)
\end{tikzpicture}
\caption{An alternative way to try fixing some issues of Algorithm \ref{fig: algoExisting}. Red nodes are failing ones, only green nodes can contribute, black messages get through, red messages do not reach its recipient. Ack-messages are dashed.}
\label{fig:complicated}
\end{figure}
The failing meter $\SM_3$ is already properly handled by Algorithm \ref{fig: algoExisting}. While $\SM_4$ was successfully contacted by $\SM_2$, it can not contribute due to the missing connection to the $\DC$. This situation is only detected but not corrected in Algorithm \ref{fig: algoExisting}. Introducing Ack-signals from the $\DC$ (Ack messages are dashed in Figure \ref{fig:complicated}) this could have been detected by $\SM_4$ and corrected by passing steps 16 and 17 of the algorithm. Note also, that for privacy reasons the first contact message of a smart meter only includes the sending list $L$ and not the sum of shares $S$. In contrast, \papertitle treats this more efficiently and elegantly by simply removing $\SM_4$ at the beginning in line 3. A case that is not even detected by Algorithm \ref{fig: algoExisting} is the missing connection of the last smart meter $\SM_6$ to the $\DC$. An alternative way to solve this problem would be that $\SM_6$ detects this (again through the missing Ack-signal on message 23 from the $\DC$) and tries to tell the smart meter before ($\SM_5$) with message 24 that it should end the round (message 25). In general it could also happen that not only the last but also the second to last smart meter has no connection to the $\DC$. This is not elegant, since the flow of information returns. Furthermore, it would require additional checks for messages that come from the ``other'' direction. While we believe that this could work, proving termination with these fixes requires more case distinctions, and this for both directions.

In terms of resilience, no solution is better than the other. In this case, resilience means the ability of the algorithm to find a path from the $\DC$ along the ring of smart meters back to the $\DC$ that includes at least $\nmin$ smart meters. More precisely: we found a configuration of network failures where \papertitle is better than this alternative, but we also found a configuration where the alternative is better. From a graph theoretic viewpoint \papertitle removes the nodes without a connection to the $\DC$ including all their links. The alternative leaves the nodes and could therefore in principle find other, longer paths. However, in order to be more resilient, more exploration of paths than shown here would be needed. A more general statement of the problem is contained in the outlook since it is out of scope of this paper. We think that it could lead to new and interesting solutions.
\end{taggedblock}
%
%
\subsection{Security and Costs}\label{sec:improveSecurity}


While Algorithm \ref{fig: algoExisting} uses homomorphic hashes to check correctness of the result, 
 in \papertitle, 
the hashes are omitted, since the measurements are assumed to be honest. This is beneficial since they are the  computationally most costly part for the smart meters. Note that we explicitly ruled out the billing use case where parties are less trusted and protocols should typically yield verifiable results \cite{Rial11a}.

Secondly, Algorithm \ref{fig: algoExisting} also has a weakness against eavesdroppers: after  eavesdropping all the messages to and from a smart meter one can compute $\tilde{m}_{i,t}-(S_{i}-S_{i-1})= m_{i,t}+s_i^0$. Since the share $s_i^0$ is static, one could easily detect changes in metering values, which are the basis for non-intrusive appliance monitoring algorithms \cite{Hart92a}. The static share $s_i^0$ is therefore substituted by a round-dependent pseudo-random number produced by a Pseudo-Random Function PRF that is computable by $\SM_i$ and the $\DC$ using pre-shared keys $k_i$ as PRF$_{k_i}(t)$.

Thirdly, Algorithm \ref{fig: algoExisting} would also yield an aggregate, if even a single household would contribute to the aggregate. Since this corresponds to a clear privacy break, this case is explicitly ruled out by only sending the information needed to recover the aggregate, if at least a minimum number of households $N_{\mathrm{min}}$ participate (line 25 of algorithm \ref{fig: algoGeneralV02}).
Finally, privacy is formally proven using game-based proofs in Section \ref{sec:privacy2}.

\subsection{Enabling Homomorphic Encryption} \label{sec:improveFlow}

Although being more costly, \papertitle can be generalized to also achieve privacy by using homomorphic encryption instead of masking. 
If homomorphic encryption is to be employed, the $\DC$ holds a secret key $sk$, and the smart meters know the corresponding public key $pk$.

Our key contribution is that the computation is decoupled from the flow of messages. Thus, the blue lines in Algorithm \ref{fig: algoGeneralV02}, representing computations, were intentionally left unspecified for proving the termination, and can be replaced easily with masking or homomorphic encryption based solutions, as we describe next.
To best of our knowledge, this algorithm is the first one that can be used with masking and homomorphic encryption, interchangeably.

%
In order to specify the privacy-preserving method (masking or homomorphic encryption), the computations in the lines 2, 6, 10 and 29 (marked blue)  must be specified using the substitutions in Table \ref{tab:computations}. There, $S=E_{pk}(m)$ denotes the homomorphic encryption of message $m$ with the public key $pk$ and $m=D_{sk}(S)$ denotes the decryption of the ciphertext with the secret key $sk$. Recall that for homomorphic encryption the decryption of the product of ciphertexts equals the sum of the plaintexts, i.e. $D_{sk}(S_1\cdot S_2)=m_1+m_2$. 
When employing homomorphic encryption, the first messages in line 2 from the smart meters to the $\DC$ seem to be unnecessary. However, these messages are also needed for the algorithmic part in order to determine $\Lcand$. Consequently, in the first message, $data_i$ is replaced by an empty value.

\begin{table*}[ht]
    \centering
    \begin{tabular}{|l|l|l|l|}
    \hline
      \textbf{Line} & \textbf{Actor}     & \textbf{Masking}  & \textbf{Homomorphic Encryption}\\
      \hline
      Line 2    & All $\SM_i$ & $ \tilde{m}_{i}= m_{i,t}+s_{i} + \PRF{j} \mod k$ and send $\langle t,i,\tilde{m}_{i}\rangle$ to $\DC$ & send $\langle t,i,\{\}\rangle$ to $\DC$.\\
      Line 6 & $\DC$ & chooses random $s_0\in \mathbb{Z}_k$ and sets $S=s_0$  & sets $S=S_0=E_{pk}(0)$\\
      Line 10    & $\SM_i$ & $S_{i}= S+s_{i} \mod k$ & $S_i=S\cdot E_{pk}(m_{i,t})$\\
       Line 29 & $\DC$   & $A_t = -S+s_0+\sum\limits_{i \in \Lact} \tilde{m}_{i} - \sum\limits_{i \in \Lact} \PRF{i} \mod k $ &  $A_t=D_{sk}(S)$\\
       \hline
    \end{tabular}
       \vspace*{0.04cm}
    \caption{Specification of the privacy-preserving computations in algorithm \ref{fig: algoGeneralV02} using either masking or homomorphic encryption}
    \label{tab:computations}
    \vspace*{-4ex}
\end{table*}



Analogously to the sum of the shares, the product of the ciphertexts of the participating smart meters measurements is now calculated. Due to the homomorphic property of the encryption system, the decryption of this product is the sum of the plaintexts, i.e. the sum of the measurements. This is formally proven in Section \ref{sec:termination}. 

\section{Guaranteed Termination and Correctness} \label{sec:termination}

\begin{theorem}[Termination, Resilience and Correctness]
	Assuming that during a single aggregation round communication channels and smart meters are either on or off and that the $\DC$ always works, the aggregation scheme shown in Algorithm \ref{fig: algoGeneralV02} terminates properly despite these errors. More precisely:
	\begin{itemize}
	    \item[(i)] The aggregation round properly finishes at the $\DC$. All smart meters are inactive after the aggregation round. $\DC$ only acts at the beginning and at the end of the round.
	    \item[(ii)] Smart meters are either never active during the round or active only once (except initialization). All active smart meters contribute to the aggregate.
	    \item[(iii)] The termination of the round is not triggered by the $\DC$, but by the smart meters. If less than $\nmin$ smart meters are available for aggregation, the $\DC$ cannot obtain the aggregate measurements.
	    \item[(iv)] The $\DC$ obtains the correct aggregate if the privacy-preserving computations are done using either masking or homomorphic encryption as shown in Table \ref{tab:computations}.
	\end{itemize}

\end{theorem}

Note that the second part of the theorem states the algorithm proceeds in a single forward pass along the sending list such that no backward steps or additional rounds are needed. The fact that all active parties contribute to the sum is notable, since it is does not stem from a fix but from the simplification to send the masked measurements during initial part (line 2).
\begin{taggedblock}{short,medium}
\textbf{Proof Sketch\footnote{Due to space restrictions, we provide all full proofs on arXiv.org.}:}
\end{taggedblock}
\begin{taggedblock}{long}
\textbf{Proof Sketch}
\end{taggedblock}
(i) and (ii): In Line 2 all smart meters send a message to the $\DC$. The $\DC$ forms an ordered list $\Lcand$ of all the smart meters from which it retrieved a message. Only these candidate smart meters may be contacted in any later step (line 15) and contribute to the aggregate. Due to the assumption about the links, the last smart meter in the round is guaranteed to be able to connect to the $\DC$ because the link has already worked in this round and deliver the information about the aggregate (line 26). 
Not all smart meters of $\Lcand$ will contribute: the ones that are not reachable from the active smart meter will be skipped (line 20).
\begin{figure}
	\centering
	\def\xSMeins{0}
\def\xSMzwei{2.5}
\def\xSMdrei{5}
\begin{tikzpicture}

\Vertex[RGB,color={127,201,127}, y=0,x=\xSMeins, label=SMi]{SM1}
\Vertex[color=white, y=0, x=1, shape=rectangle, label=DC]{DC1}
\Edge [RGB,color={127,201,127}](SM1)(DC1)

\Vertex[RGB,color={127,201,127}, y=0,x=\xSMzwei, label=SMi]{SM2}
\Vertex[color=white, y=0, x=3.5, shape=rectangle, label=DC]{DC2}
\Vertex[RGB,color={127,201,127}, y=-1,x=\xSMzwei, label=SMj]{SM2next}
\Edge [RGB,color={127,201,127}](SM2)(DC2)
\Edge [RGB, color={127,201,127}](SM2)(SM2next)

\Vertex[RGB,color={127,201,127}, y=0,x=\xSMdrei, label=SMi]{SM3}
\Vertex[color=white, y=0, x=6, shape=rectangle, label=DC]{DC3}
\Vertex[color=white, y=-1,x=\xSMdrei, label=SMj]{SM3next}
\Edge [RGB,color={127,201,127}](SM3)(DC3)
\Edge [opacity=.5, style=dashed](SM3)(SM3next)






\end{tikzpicture}
\caption{Cases that occur in the proof (filled: contributing smart meters): Left (C1): active smart meter is the last one. Middle (C2)/right (C3): active smart meter is not the last, the connection to the next smart meter is working/not working.}
\label{fig:proofCases}
\vspace*{-3ex}
\end{figure}
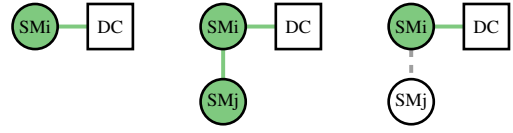

The most difficult part is the proving proper termination and proving inactiveness of all smart meters after the round. This part needs a distinction into 3 cases (see Figure \ref{fig:proofCases}): C1 is the standard end case: the next party is the $\DC$ (with guaranteed working link as stated above). C2 is the standard case during the round: the next party is a smart meter and the connection to it works. C3 is the case where error-resilience is needed: the next party is a smart meter and the connection to it does not work. This unreachable smart meter is removed from $\Lcand$ (line 20) and the case is then further split into 2 subcases, depending on the distinction, if the still active smart meter is the last one (case C3.1) or not (case C3.2). While in cases C1 and C3.1 the round terminates with the DC, in cases C2 and C3.2 the list $\Lcand$ gets shorter, which must lead to an empty list at some later step. In C1, C2 and C3.1, the smart meter can easily be seen to be inactive  after  the  round.  In  case  C3.2, the  smart  meter  stays active  and  only  gets  passive  when  it  leaves  the  while  loop (tryNext=false)  through  case  C2  (line18)  or  C3.1  (line23) in a later iteration.

(iii): Termination is triggered by the active smart meter in lines 12 and 21 when either no smart meter is left or the minimum number of smart meters is not achievable (procedure checkLast). The check for the minimum number of smart meters is done by the last smart meter in line 2 of procedure sendFinalMessage.

(iv): 
Statement (ii) ensures that $S$ contains exactly the data of the set $\Lact$ of active smart meters.

Masking: the $\DC$ gets all the masked values $\tilde{m}_{i}= m_{i,t}+s_{i} + \PRF{i} \mod k$ in line 2. The correctness then follows from inserting this in $A_t$ and modular arithmetic.

Homomorphic variant: the calculations in lines 6 and 10 lead to  $S=E_{pk}(0)\cdot \prod\limits_{i \in \Lact} 
\cdot E_{pk}(m_{i,t})$. Correctness follows from insertion in $A_t$ and the homomorphic property. \hfill $\qed$

\begin{taggedblock}{long}

\textbf{Full Proof:}\\
\textbf{(i)-(iii)}: The most intricate part is to prove statement (i), the other two statements are easier and will be proven along the way. 
The proof is done by going through the whole algorithm and all possible distinct cases. As the algorithm this part is split into the first part where mainly the $\DC$ acts (A1: lines 1-8), the flow along the smart meters (A2: lines 9-27) and the finish at the $\DC$ (A3: lines 28-end).

\textbf{(A1)} Part $\DC \to \SM_i$: In Line 2 all smart meters send their initial message to the $\DC$. The $\DC$ forms an ordered list of all the smart meters from which it retrieved the message. This can be done since the $\DC$ is assumed to function the whole time. The same argumentation holds for all computations that the $\DC$ does later (lines 3-7, line 31). This list $\Lcand$ includes the \textit{remaining} candidate smart meters that may contribute to the aggregate. For all these smart meters in this list, due to the assumptions of the theorem the following holds: (w1) since they worked with sending, the smart meters will work also in all subsequent steps of this round;  (w2) similarly, also the links from each of these smart meters to the $\DC$ will work for the rest of the round. The other smart meters will not occur any more in the round, since only these remaining smart meters in the set $\Lcand$ will receive messages. This is one set of smart meters that will never be active in statement (ii).
In line 4, the $\DC$ checks already, if enough smart meters remain to contribute the the aggregate. While this is not needed for proving (iii), an early stop would save resources. In line 5, since no smart meter is active yet, the $\DC$ sets the set of active smart meters to be empty. Then it chooses the first smart meter in from the ordered list (line 7) and sends the message with the data ($S$) to it. Note that the message is guaranteed to get through because of (w2). The message not only provides information, it also activates the contacted smart meter which takes over. Note that the contacted smart meter is guaranteed to work because of (w1). The $\DC$ will only get active in the final aggregation step. Thus, the part $\DC \to \SM_i$ is finished.

\textbf{(A2)} Part $\SM_i \to \SM_j$: The smart meter first acknowledges the receipt of the message to its preceding, activating party which can then get inactive (see case C2 later). Note that this message gets through because of (w2), also the activated smart meter will continue to work due to (w1). After calculations in line 10, the active smart meter updates the set of active smart meters. More importantly for finishing of the algorithm, it removes itself from the list $\Lcand$ of remaining smart meter which therefore gets shorter with each active smart meter (w3). Additionally, since only smart meters on this list are contacted later on (line 15), it can never be activated again in this round  (statement ii).

Now, three cases need to be distinguished (see Figure \ref{fig:proofCases}): C1) is the standard end case: the next party is the $\DC$ (with guaranteed link due to (w1)). C2) is the standard case: the next party is a smart meter and the connection to it works. C3) is the case where error-resilience is needed. The next party is a smart meter and the connection to it does not work. Note that for all 3 cases the connection to $\DC$ works due to the simplification done in line 3. 


\textbf{Case C1)}: In line 12, the smart meter checks, if it is the last smart meter (no smart meter is remaining, thus $\Lcand=\{\}$). Since this is true, $tryNext$ is false and the while loop is skipped as a whole, then the smart meter sends the final message in line 28 and gets inactive. Due to (w2) this message gets through and the round proceeds to part A3, where it is finished at the $\DC$. This proves termination of part A2 for this case, the inactivity of this smart meter for statement (i). Note also that the termination of part A2 was triggered by the smart meter. The $\DC$ obtains the information $S$ that is needed to calculate the aggregate in part A3 from a message that has been sent by this smart meter in Procedure sendFinalMessage. The condition that the number of contributing smart meters exceeds $\nmin$ is already built into that procedure sending an empty $S$ if $\vert \Lcand \cup \Lact \vert < \nmin$ which forces $\vert \Lact \vert \leq \vert \Lcand \cup \Lact \vert < \nmin$  (statement (iii)).

\textbf{Case C2)}: The link to the next smart meter in the list works. Therefore other smart meters are remaining $\Lcand \neq \{\}$ and the while loop is entered. There, the message in line 16 gets through. Due to (w1) and (w2) the contacted smart meter $j$ successfully acknowledges and gets active. Smart meter $i$ gets out of the loop by line 18 and gets inactive (i). The final message is not sent since $isLast$ was false and is not changed in the while loop. This case thus results in a new active smart meter with a shorter list $\Lcand$.

\textbf{Case C3):} this is the most challenging case. It arises due to a non-working link: also here other smart meters are remaining ($\Lcand \neq \{\}$) and the while loop is entered. Since the connection to the next smart meter does not work, the active smart meter gets no $\Ackring$ and stays active entering the else-part of the while loop. It removes the next smart meter from $\Lcand$ which gets shorter. Again, since only smart meters from $\Lcand$ are contacted, the unsuccessfully contacted smart meter will not be activated later in this round although it has successfully sent the message to the $\DC$ in part A1. Then the active smart meter checks, if it is the last smart meter after the removal.\\ 
If yes (\textbf{Case C3.1}), it escapes the while loop, sends the final message and gets inactive as in case C1. So part A2 finishes (triggered by the smart meter with the final message, statement (iii)) and the round is then finished at the $\DC$. Note that in this configuration of working or failing links the case C1 will not occur, the transition to part A3 happens through this sub-case.\\ 
If the active smart meter is not the last one (\textbf{Case C3.2}), it stays in the while loop and contacts the next smart meter. So the case distinction starts again with cases C2 and C3.1 being possible again. However, in contrast to before now the list $\Lcand$ is shorter by one entry.\\
Summarizing: in cases C1 and C3.1 part A2 terminates and the algorithm proceeds with the $\DC$ in part A3. In both, cases C2 and C3.2, the list $\Lcand$ gets shorter. This can not happen forever, at a later iteration the list gets empty and this part A2 algorithm must therefore finish (with either case C1 or C3.1). In C1, C2 and C3.1, the smart meter is already shown to be inactive after the round. In case C3.2 the smart meter stays active and only gets passive when it leaves the while loop ($tryNext=false$) through case C2 (line18) or C3.1 (line 23) in a later iteration as already shown above.
Note that in all these cases only the smart meter acted, the $\DC$ only gets active at the transition to part A3.

\textbf{(A3)} Finish at $\DC$: since in A3 only the $DC$ acts no smart meters are activated there and due to the assumption that the $DC$ always works part A3 will terminate. It may be that it can not compute the aggregate, if it does not have the data needed for it, but here only the termination is proven.


\end{taggedblock}
\begin{taggedblock}{long}
\textbf{(iv)}: 
Now the algorithm is analyzed with specified to privacy preserving computations. Due to (ii) only smart meters that get active will provide information through $S$ such that they contribute to the aggregate, inactive smart meters do not contribute. Statement (iii) also already states that if less than $\nmin$ smart meters have been active only the information about the end of the round is transferred in the final message. In this case the $\DC$ does not have the needed information to compute the aggregate.
Therefore correctness of the aggregate means that the aggregate for the round obtained is the exact sum of the measurements of all the active smart meters that were active if they were at least $\nmin$. 
\[
A_t=\sum_{i \in \Lact}m_{i,t}, \quad \mathrm{if} \ | \Lact|\geq \nmin
\]
Note that honest parties follow the protocol therefore the protocol is executed as in Algorithm \ref{fig: algoGeneralV02}.

Masking: in line 2, the $\DC$ gets the masked values of all smart meters in $\Lcand$. Since all active smart meters are in $\Lcand$ (lines 7 and 15, respectively), it consequently has the masked values $\tilde{m}_{i}= m_{i,t}+s_{i} + \PRF{j} \mod k$ of all active smart meters. Since in line 10 only active smart meters act,  $S=s_0+\sum\limits_{i \in \Lact} s_i$ due to the corresponding substitutions in lines 6 and line 10. 
As shown in in the previous theorem, only if at least $\nmin$ smart meters have been active, this $S$ is sent to the $\DC$ in line 26. The correctness then follows from inserting all this in $A_t$ and modular arithmetics which allows the summations to be done in arbitrary order. It can be then seen that all the terms, i.e. $s_0  \mod k$, $\sum\limits_{i \in \Lact} s_i  \mod k$ and $\sum\limits_{i \in \Lact} \PRF{i} \mod k$, cancel themselves out and only the sum of the measurements $\sum\limits_{i \in \Lact} m_{i,t} \mod k$ remains.
\begin{eqnarray*}
      A_t & = & -S+s_0+\sum\limits_{i \in \Lact} \tilde{m}_{i} - \sum\limits_{i \in \Lact} \PRF{i} \mod k\\
      & = & -(s_0+\sum\limits_{i \in \Lact} s_i)+s_0\\
     &  & +\sum\limits_{i \in \Lact} \left( m_{i,t}+s_{i} +\PRF{i} \mod k\right) \\
     & &  - \sum\limits_{i \in \Lact} \PRF{i} \mod k\\
 \end{eqnarray*}
 Note that $k$ must be chosen bigger than $\sum\limits_{i \in \Lact} m_{i,t}$ as already stated in \cite{Knirsch16a}.

Homomorphic encryption: in line 2, the $\DC$ gets the indices of all smart meters so that it can determine $\Lcand$ needed for the algorithm. Analogously to masking, through the calculations in line lines 6 and 10 $S$ is now the product $E_{pk}(0)\cdot \prod\limits_{i \in \Lact} 
\cdot E_{pk}(m_{i,t})$. The correctness follows from inserting this in $A_t$ and the homomorphic property of the encryption scheme and that the fact that  $\DC$ (only the $\DC$) has the secret key 
\begin{eqnarray*}
D_{sk}(S)&=&D_{sk}\left(E_{pk}(0)\cdot \prod_{i \in \Lact}S_i\right)\\
&=&D_{sk}\left(E_{pk}(0)\cdot \prod_{i \in \Lact}E_{pk}(m_{i,t})\right)=\sum_{i \in \Lact}m_{i,t}\\
\end{eqnarray*} 
\hfill $\qed$
\end{taggedblock}

\section{Computational and Network Costs} \label{sec:costs}
Smart meters have limited computing power: in one round of the protocol that uses masking, a smart meter only needs to perform 3 summation operations and the generation of a pseudorandom number, and the normal accompanying security measures only use light symmetric-key operations. When using the asymmetric homomorphic encryption, the encryption of the measured value dominates the needed computation power, which is more than needed for masking.

Communication complexity is also low, since the algorithm proceeds forward only: compared to a privacy-less solution, two additional messages per smart meter is our overhead, on average: in addition to sending the (masked) metering value, on average one Ack-message backwards and one message forwards along the ring is needed. 
The latency until the aggregate is available is $\mathcal{O}(N\Delta T)$.

\section{Privacy Guarantee  and Proof} \label{sec:privacy2}
Privacy of smart meters in an aggregation protocol is preserved as long as no adversary is able to infer anything about the individual measurements. However, note that privacy is in any case limited since the protocol's task is to provide the $\DC$ with the aggregate value. 
Regardless of the underlying protocol, the aggregate itself may allow inference of personal information if the aggregate consists of too few values. While we introduced $\nmin$ for this reason, data analysis attacks on whole load profiles such as the ones described in \cite{Buescher17a} are explicitly not considered in this section.

\subsection{Assumptions and Game-Based Privacy Definition}

The privacy objective will be captured by defining a privacy game, extending the smart meter’s data unlinkability game defined in \cite{Unterweger18b}. The game is played between an adversary and a challenger. The adversary controls all colluding adversarial entities, whereas the challenger acts on behalf of the remaining honest parties.

\textbf{Adversarial Model:}
As in \cite{Unterweger18b}, attackers are considered as honest-but-curious entities that follow the protocol but can collude and share their information to infer information about others. The only restriction in the game is that at least two honest smart meters (out of $N$ total smart meters) are out of the adversary's control. These smart meters are called \textit{challenged smart meters}. Note again, that communication 
links between smart meters and the $\DC$ are assumed secure and authenticated. 

In this paper, the adversary's power is extended  by granting some control over network failures: the bidirectional links can be turned off and on but no such changes are allowed during \textit{one round} of the aggregation protocol. Note that links are only cut, no messages can be observed.
This cutoff of communication lines could happen during a network attack even without touching smart meters. While this attack is likely having a different goal, privacy will be preserved as long as the two challenged smart meters remain contributing.
More formally, the adversary specifies the network failures by outputting a sub-graph $G'=(V', E')$ of $G$ with the same parties $V'=V$ but only a subset of working links $E'\subseteq E$.
Throughout the game, messages can only be exchanged between two entities $V_i$ and $V_j$ when the connecting edge is in $E'$ (i.e., when $(V_i,V_j) \in E'$, solid lines in Figure \ref{fig:exNetwork0}). Otherwise, the link is off (dashed lines) and no data transmission can take place.

Second, the adversary is now allowed to obtain more information. In step 2 of the extended game, the adversary additionally knows the individual measurements of any subset of the $N-2$  \textit{non-challenged} smart meters. This additional power models a scenario where that knowledge might get exposed to the adversary indirectly, without the households' intention. Note that this does not include the secret shares or ciphertexts, which are only known for the \textit{colluding} smart meters that are under his control.
As in \cite{Unterweger18b}, the adversary also knows the measurements of the two \textit{challenged} smart meters, he can even choose the two measurements. However, these two measurements are randomly assigned to these two honest smart meters by the challenger and the adversaries goal is to determine this assignment. The aggregation protocol preserves data unlinkability if this can only be inferred with a probability \textit{negligibly} higher than for a random guess. 

\textbf{Smart meters data unlinkability under chosen bi-directional network failure model $\Unlink$}:
\begin{enumerate}
    \item The initialization is run. The challenger gives the network graph $G=(V, E)$ to the adversary. 
    \item The adversary $\A$ outputs a pair of measurements $m_0, m_1$  within the measurement domain to be randomly assigned to two smart meters $SM_{i^*}$ and $SM_{j^*}$ (the challenged smart meters). The adversary also specifies the measurements of all the remaining smart meters, i.e., $MList=\{(SM_i, m_i)\}$ for $i\notin \{ i^{*},j^{*}\}$. The adversary sends out the network failure model $G'=(V', E')$ as well as an ordered list $L=(DC, SM_1, SM_2,\cdots, SM_N, DC)$ that indicates the arrangements of smart meters in the aggregation protocol.
    \item The challenger checks that $L$ starts and ends with $DC$, and contains all the smart meters exactly once. The challenger also checks that the challenged smart meters $SM_{i^*}$ and $SM_{j^*}$ can contribute to the aggregation in the network failure model $G'=(V', E')$ provided by the adversary, and that it is a valid failure model ($V' = V \wedge E' \subseteq E$). If any of the checks fail, the challenger stops and the adversary loses the game with the game’s output being $0$. Otherwise (if all checks succeed), the challenger assigns the measurements of smart meters according to $MList$. For the  challenged smart meters, he draws a random bit $b \in \{ 0, 1 \}$ and assigns $m_b$ to $SM_{i^*}$ and $m_{\hat{b}}$ to $SM_ {j^*}$.
    \item The challenger and the adversary proceed with the steps of the aggregation protocol using the network failure model provided by the adversary. 
    \item The adversary outputs her guess $b^{'}$ to the challenger.
    If $b = b^{'}$, then the adversary wins, and the output of the game is set to $1$, Otherwise, the adversary loses and the game’s output is $0$.
\end{enumerate}

\begin{definition}[Data Unlinkability] \label{theorem:dataUnlink}
An aggregation protocol provides smart meters’ data unlinkability under chosen bi-directional network failure model, if for every probabilistic polynomial time (PPT) adversary $\A$, the success probability in $\Unlink$ is negligibly higher than a random guess, i.e.
\begin{align}
    Pr[ \Unlink = 1] \le \frac{1}{2}+ negl(\lambda) 
\end{align}
for some negligible probability $negl(\lambda)$, where $\lambda$ is the security parameter.
\end{definition}

\textbf{Maximal collusion sets:}
The privacy guarantee lists the maximum possible corruption of various parties where privacy of the uncorrupted smart meters is still preserved. Privacy would be broken if either any more party is added or any kind of corruption is increased. Therefore, the more allowed corruption the better. 

\begin{figure}[htb]
    \centering
    \begin{minipage}{0.3\textwidth}
    
      \begin{collusiontable}[baseline=(current bounding box.base)]
        \entityname{$\SM\ {\scriptstyle 1..N}$}
        \entityname{$\DC$}
        \nomoreentitynames

        \collusion[0.95]{informationtheoretic} \collusioncomment[white]{left}{\tiny $N-2$}
        \collusion{insecure}
        \nextcollusionset

        \collusion[0.95]{hard} \collusioncomment[white]{left}{\tiny $N-2$}
        \collusion{informationtheoretic}
        \nextcollusionset
      \end{collusiontable}
      
    \end{minipage}
          \hfill
  \begin{minipage}{0.12\textwidth}
    \centering
    \printcolorkey[baseline=(current bounding box.base)]
  \end{minipage}
        
    \caption{Visualization of privacy guarantee when masking is used as a privacy-preserving method. Black: Fully corrupted party: all internal parameters like secret shares or keys are observed. Gray: metering values are observed,  communication links can be turned off (visualized using the tool developed in \cite{Unterweger18b}). The guarantee is information-theoretic and computationally-hard for sets 1 and 2, respectively.}
    \label{fig:privVisu1}
    \vspace*{-3ex}
\end{figure}
 
\textbf{Game based privacy proofs:}
Privacy with respect to the maximal collusion sets is proved using game-based proofs as described in \cite{Unterweger18b}. 
A game-based proof is formulated as two nested games. The inner game is the game of breaking privacy stated as an unlinkability game, whereas the outer game is a common problem that is known to be \textit{(computationally) hard}. The security proof embodies a valid reduction from the inner game to the outer game  indicating that if one manages to break privacy in the inner game, then one can also win the outer game.  Since winning the outer game is known to be hard, so is the inner game.
\begin{taggedblock}{long}
\subsection{IND-CPA and PRG indistinguishability games}
This section supplies common security definitions underlying the security proofs in the subsequent subsections.

\textbf{Negligible function: } \label{footnote:negl}
A function $f$ is negligible \cite{katz2020introduction} 
 :$\Leftrightarrow \forall \ \mathrm{polynomial \ functions} \ p \ \exists N\in \mathbb{N}: \ f(n)<\frac{1}{p(n)}$.

\textbf{Indistinguishable Pseudo Random Number Generator:} A pseudo random number generator PRG \cite{blum1984generate} is a function whose output should be indistinguishable from a true random number, given a random seed. More formally, let $l$ be a polynomial function with $l(m)>m$, $\lambda$ denote the security parameter and  $y\leftarrow \{0,1\}^{l(\lambda)}$ be a true random number. Let $PRG$ be a function $PRG: \{0,1\}^{\lambda} \rightarrow\{0,1\}^{l(\lambda)}$, and $s\leftarrow \{0,1\}^{\lambda}$ be a random seed and $x=PRG(s)$ be its output. Now consider a game between a challenger $\mathcal{C}_{PRG}$ and a distinguisher $D$ (here the adversary is typically called distinguisher). $\mathcal{C}_{PRG}$ randomly chooses $b_{PRG}\in\{0,1\}$ and presents $x$ to $D$, if $b_{PRG}=1$ and $y$ otherwise. $PRG$ is then a pseudo random number generator, if for every probabilistic polynomial time (PPT) distinguisher $D:\{0,1\}^{l(\lambda)}\rightarrow\{0,1\}$ there exists a negligible function $negl$ such that
\[
|\Pr[D(x)=1]- \Pr[D(y)=1]|= negl(\lambda)    
\]
The above definition can be extended to  indistinguishability of multiple samples comparing sequence (vector) of outputs $\vec{x}=(x_1,\cdots,x_n)\leftarrow PRG(s)$ and sequence (vector) of true random numbers $\vec{y}=(y_1,\cdots,y_n)$
\begin{equation}\label{eq:vectorPRG}
|\Pr[D(\vec{x})=1]- \Pr[D(\vec{y})=1]|= negl(\lambda)    
\end{equation}

\textbf{Indistinguishability under Chosen Plain-text Attack (IND-CPA):} A public key encryption scheme $\pi$ is said to be IND-CPA secure, if the encryption of two arbitrary messages cannot be distinguished. This is formulated as a game between a challenger $\mathcal{C}$ and an adversary $\mathcal{A}$, where $\mathcal{A}$ performs only negligibly better than random guessing. More formally, consider a public key encryption scheme $\pi = (KeyGen, Enc, Dec)$ and denote with $(pk,sk)\leftarrow KeyGen(1^\lambda)$ the generated public and secret keys.
Let $\mathcal{A}$ be a PPT adversary that chooses two arbitrary messages $(m_0,m_1) \leftarrow \mathcal{A}(pk)$. The challenger $C$ (who is the only party knowing $sk$) randomly selects one of these messages and encrypts it
\[
b\leftarrow\{0,1\}; \\ c \leftarrow Enc_{pk}(m_b).
\]
The adversary $\mathcal{A}$ is given the ciphertext $c$ and tries to guess $b$, i.e., if it is given the encryption of $m_0$ (case $b=0$) or the encryption of $m_1$ (case $b=1$)
\[
b'=\mathcal{A}(c).
\]
The public key encryption scheme $\pi$ is 
IND-CPA secure if for all PPR $\mathcal{A}$, there exists a negligible function $negl(.)$ such that  
\begin{equation} \label{eq:probINDCPA}
Pr[ b=b' ] \le \frac{1}{2} + negl(\lambda)
\end{equation}
As before, the definition can be extended to indistinguishability of multiple encryptions substituting $m_0$ with $\vec{m_0}=(m^1_0,\cdots,m^n_0)$, and $m_1$ with $\vec{m_1}=(m^1_1,\cdots,m^n_1)$ and $c$ with $\vec{c}=(Enc_{pk}(m^1_b),\cdots, Enc_{pk}(m^n_b))$ (the challenger still picks only one bit $b$).


\end{taggedblock}

\subsection{ Privacy Analysis of the Masking Aggregation Protocol}
The following theorem \ref{thm_masking} states  the privacy guarantee of the protocol when masking is used: essentially, it states that the two maximal collusion sets are the ones described in Figure \ref{fig:privVisu1}. 
\begin{theorem}[Privacy of the aggregation protocol using masking] \label{thm_masking}
Consider the protocol \papertitle of Algorithm \ref{fig: algoGeneralV02} with privacy-preserving computations done using masking as described in Table \ref{tab:computations}. The following holds where privacy is as defined in the $\Unlink$ game:
\begin{itemize}
    \item[(i)] The protocol provides information-theoretic privacy against an adversary controlling $N-2$ smart meters. 
    \item[(ii)] If the underlying Pseudo-Random number Generator PRG (used for producing masking shares) is cryptographically secure (as defined in \cite{katz2020introduction}), the protocol based on masking provides privacy against an adversary controlling $\DC$.
    \item[(iii)] Both collusion sets above are maximal.
\end{itemize}
\end{theorem}

\subsubsection{Proof of Theorem \ref{thm_masking}.(i)}
The only information about the two challenged smart meters an adversary  $\A$ controlling $N-2$ smart meters gets are the masking sums $S$ embedding no measurements. Since the masked measurements are only submitted to the $DC$ controlled by the challenger, they remain unknown to  $\A$. As $\A$'s view is independent of the challenged smart meters measurements (hence independent of $b$),  unlinkability is preserved information-theoretically. 
Following the preceding reasoning, the masking protocol preserves information-theoretic privacy in an extended  adversarial set consisting of $N-1$ smart meters where the privacy is defined as the inability of the adversary in distinguishing between two arbitrary measurements ($m_0$ and $m_1$) of the non-colluding smart meter. 

\subsubsection{Proof of Theorem \ref{thm_masking}.(ii)}
We show that if there exists a PPT adversary $\A$ which can win $\Unlink$, then we can construct a PPT adversary $B$ who can win the indistinguishability game of PRG. 

\begin{taggedblock}{long}
\textbf{Proof Overview: }
\end{taggedblock}
In the reduction proof, $B$ acts as the challenger for the unlinkability game and at the same time as the adversary for the PRG indistinguishability game. $B$ receives a vector of two values $\overrightarrow{S}=<s_0,s_1>$ which are either truly random or generated by a pseudo-random number generator. $B$ leverages the success probability of $\A$ to win the PRG game by using $s_0,s_1$ to mask the measurements of $SM_{i^*}$ and $SM_{j^*}$ (the challenged smart meters).
At the end of the game, $\A$'s view toward the challenged smart meters is limited to $S=s_0+s_1+\sum_{i \in \{SM_i\in \Lact / \{SM_{i^{*}},SM_{j^{*}}\} \}}$ and their individual masked measurements i.e., $\tilde{m}_{i^{*}}$,  $\tilde{m}_{j^{*}}$. 
The value $S$ is independent of $b$ and no help for $\A$. In order for the adversary to make a correct guess, it needs to exploit $\tilde{m}_{i^{*}}$ and $\tilde{m}_{j^{*}}$. If $\overrightarrow{S}$ is truly random, then $\tilde{m}_{i^*}$ and $\tilde{m}_{j^*}$ look completely random to $\A$, which gives $\A$ no advantage to win the unlinkability game. In case  $\overrightarrow{S}$ is pseudo-randomly generated, $\A$ may have some advantage $\epsilon(\lambda)$ to win the unlinkability game.
Note that $\epsilon(\lambda)$ tightly relates to the distinguishability of the PRG from truly random number generators: if $\A$'s output $b'$ equals to $b$, then $B$ returns $0$ to the PRG challenger signifying that $ \overrightarrow{S}$ is pseudo-random, otherwise, $B$ outputs $1$ in the PRG game meaning $\overrightarrow{S}$ is truly random. However, as the deployed PRG is secure, $\epsilon(\lambda)$ is a negligible value which means the $\A$'s advantage in winning the unlinkability game is also negligible. 
\begin{taggedblock}{long}

Figure \ref{fig:distinguisher} visualizes the  reduction proof and shows how winning the unlinkability game $\Unlink$ can be used to win the PRG indistinguishability game (which is known to be computationally hard).
\begin{figure*}
	\centering
	\includegraphics{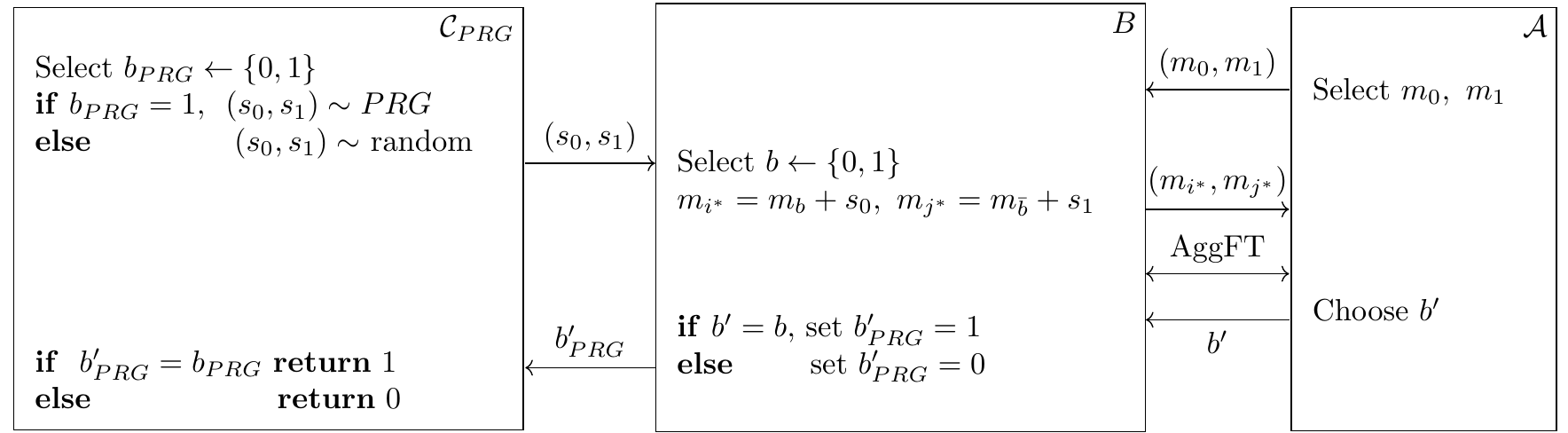}
		\caption{Construction of a PPT distinguisher $B$ which plays the unlinkability game as the challenger with the adversary $\A$ and which is the attacker of the indistinguishability game. If $\A$ wins the unlinkability game, $B$ can distinguish pseudorandom from random numbers. However, this is computationally hard.}
\label{fig:distinguisher}
\end{figure*}
\textbf{Formal proof}
\begin{enumerate}
    \item The initialization part of the aggregation protocol is run as explained next. $B$ plays on behalf of  all the smart meters whereas $\A$ has the role of $DC$. $B$ is given a vector of two values $\overrightarrow{S} =<s_0,s_1>$ from the outside PRG indistinguishability game.  $B$ gives the network graph $G=(V, E)$ to  $\A$.  $\A$, playing as $DC$, outputs the $PRF$ keys $k_1,\cdots,k_N$ corresponding to $SM_1,\cdots,SM_N$, respectively, thereby finalizing initialization.
    \item $\A$ outputs a pair of measurements $m_0, m_1$  within the measurement domain. $\A$ also determines two smart meters $SM_{i^*}$ and $SM_{j^*}$ (the challenged smart meters) to which the given measurements shall be randomly assigned.  $\A$ also specifies the measurements of all the remaining smart meters, i.e., $MList=\{(SM_i, m_i)\}$ for $i\notin \{ i^{*},j^{*}\}$.  $\A$  sends out the network failure model $G'=(V', E')$ as well as an ordered list $L=(DC, SM_1, SM_2,\cdots, SM_N, DC)$ that indicates the arrangements of smart meters in the aggregation protocol.
    \item  $B$ checks that $L$ starts and ends with $DC$, and contains all the smart meters exactly once.  $B$ also checks that the challenged smart meters $SM_{i^*}$ and $SM_{j^*}$ can contribute to the aggregation in the network failure model $G'=(V', E')$ provided by $\A$, and that it is a valid failure model ($V' = V \wedge E' \subseteq E$). If any of the checks fail,   $B$ aborts the game. Otherwise (if all checks succeed),  $B$ assigns the measurements of smart meters according to $MList$. 
    For the  challenged smart meters, a random bit $b \in \{ 0, 1 \}$ is drawn and $m_b$ is assigned to $SM_{i^*}$ and $m_{\hat{b}}$ to $SM_ {j^*}$.
    \item  $B$ and $\A$ proceed with the steps of the aggregation protocol using the network failure model provided by  $\A$. 
    \begin{enumerate}
         \item $B$ attempts the execution of line 2 of the aggregation. For each $SM_i$ $i\in\{1,\cdots,N\}$, $B$ checks whether $(SM_i,DC) \in E'$ and sends the masked measurements $\tilde{m_i}=m_i+s_i+\PRF{i}$  to $\A$ accordingly. $B$ generates random masking factors i.e., $s_i$ for the non-challenged smart meters whereas for $SM_{i^*}$ or $SM_{j^*}$, $B$ uses $s_0$ and $s_1$ to mask their measurements as shown below, \begin{align}
            \tilde{m_{i^*}}= m_{i^{*},t}+s_{0} + \PRF{i^{*}} \mod k \nonumber \\
            \tilde{m_{j^*}}= m_{j^{*},t}+s_{1} + \PRF{j^{*}} \mod k 
        \end{align}
        \item $DC$ outputs $S$ alongside with $\Lcand$, and $\Lact$ to the first smart meter in $\Lcand$ (lines 6-9 of Algorithm \ref{fig: algoGeneralV02}). 
        $B$ sends an $\Ackring$ to $\A$.
        \item For each smart meter in  $\Lcand$, $B$ proceeds according to lines 10-27 of Algorithm \ref{fig: algoGeneralV02}. During this part, $SM_i$ (i.e., $B$) needs to forward its calculated $S_i$ to the next reachable smart meter $SM_j\in \Lcand$ where $j>i$, as such, $B$ checks whether the two smart meters have an active connection link or an edge in $E'$ i.e., $(SM_i,SM_j)\in E'$. The presence/absence of such edge corresponds to the receipt/lack of receipt of an Ack from $SM_j$ (lines 17 and 19 of Algorithm \ref{fig: algoGeneralV02}), respectively. For smart meters with non-working links, $B$ removes them from $\Lcand$. $B$ continues with the rest of the protocol. Finally, as the last reachable smart meter, $B$ runs the $sendFinalMessage()$ (lines 1-6) and sends $S$ and $\Lact$ to $\A$.
    \end{enumerate}
    \item  $\A$ outputs her guess $b^{'}$ to $B$. 
    $B$ checks whether $b==b'$, if yes, $B$ outputs 1 to the PRG indistinguishability game challenger (indicating that $s_0,s_1$ are distinguishable from truly random values), otherwise 0 (indicating that $s_0,s_1$ are indistinguishable from truly random values).
\end{enumerate}
At the end of the game, $\A$'s view toward the challenged smart meters is limited to $S=s_0+s_1+\sum_{i \in \Lact / \{SM_{i^{*}},SM_{j^{*}}\} }s_i$ and their individual masked measurements i.e., $\tilde{m}_{i^{*}}$,  $\tilde{m}_{j^{*}}$.
The value of $S$ is independent of $b$ thus makes no help for  $\A$. In order for  $\A$ to be able to make a correct guess, it needs to exploit $\tilde{m}_{i^{*}}$,  $\tilde{m}_{j^{*}}$. If $\overrightarrow{S}$ is truly random, then $\tilde{m}_{i^*}$ and $\tilde{m}_{j^*}$ look completely random to $\A$, which leaves $\A$ with no advantage to win the unlinkability game. In case  $\overrightarrow{S}$ is pseudo-randomly generated, $\A$ may have some advantage $\epsilon(\lambda)$ to win the unlinkability game. 
However, $\epsilon(\lambda)$ tightly relates to the distinguishability of the PRG from truly random number generators in (\ref{eq:vectorPRG}). 
With $R$ and $PR$ standing for \textit{Random} and \textit{Pseudo-Random}, respectively, this relation is shown in (\ref{eq:proof_masking_dc}).  Note that the reduction is constructed in such a way (in part 5 above) that the first probability is the probability to win the unlinkability game, and the second probability is 1/2 since the values are truly random.
\begin{align} \label{eq:proof_masking_dc}
    |Pr&[B \text{ outputs 1} | \overrightarrow{S} \text{ is PR}]-Pr[B \text{ outputs 1}| \overrightarrow{S} \text{ is R} ]| \nonumber  \\
    & = |Pr[b=b' | \overrightarrow{S} \text{ is PR}] - Pr[b=b' | \overrightarrow{S} \text{ is R}]|  \nonumber \\
    & = |[\frac{1}{2} +\epsilon(\lambda)] -\frac{1}{2}| = \epsilon(\lambda)
\end{align}
If the PRG is cryptographically secure, comparing (\ref{eq:proof_masking_dc}) with (\ref{eq:vectorPRG}) shows that the advantage in winning the unlinkability game $\epsilon(\lambda)$ must be negligible.

\hfill $\qed$

\end{taggedblock}

\subsubsection{Proof of Theorem \ref{thm_masking}.(iii)}
\begin{taggedblock}{short}
To prove that a collusion set is maximal, we show that any super set of the maximal set can launch a successful attack and break $\Unlink$. 
The collusion set in Theorem \ref{thm_masking}.(i) can only be extended by adding the $\DC$. With the following attack, the adversary learns the measurement of $\SM_{i^*}$: first, the adversary arranges the sending list in a way that the first active smart meter in the aggregation phase is $\SM_{i^*}$ and its next is $\SM_{k}$, i.e. $L=\{i^*,k,\ldots\}$. Since $SM_{k}$ gets $S_{i^*}=s_0+s_{i^*}$ and $\DC$ gets $\tilde{m_{i^*}}=m_{i^*}+s_{i^*}+\PRF{i^*}$ from $\SM_{i^*}$, together they recover 
$m_{i^*}=\tilde{m}_{i^*}-(S_{i^*}-s_{0}) -\PRF{i^*} \mod k$. Having learned $m_{i^*}$, $\A$ easily wins the game. 
In collusion set (ii), the adversarial set is augmented by adding one smart meter, namely $\SM_{k}$, so $\A$ can use the same attack to break privacy. \qed
\end{taggedblock}
\begin{taggedblock}{long}
To prove that a collusion set is maximal, we show that any super set of the maximal set can launch a successful attack and break $\Unlink$. \\
The collusion set in Theorem \ref{thm_masking}.(i) can only be extended by adding the $\DC$. Consider an adversary who wants to learn the measurement of $\SM_{i^*}$. The adversary arranges the sending list in a way that the first active smart meter in the aggregation phase is $\SM_{i^*}$ and its next is $SM_{k}$, i.e. $L=\{i^*,k,\ldots\}$. Moreover, the adversary outputs the network failure model $G'=(V,E')$ in such a way that there is an edge between $i^*$ and $k$ i.e., $(i^*,k)\in E'$. Since $SM_{k}$ gets $S_{i^*}=s_0+s_{i^*}$ and $\DC$ gets $\tilde{m_{i^*}}=m_{i^*}+s_{i^*}+\PRF{i^*}$ from $\SM_{i^*}$, together they recover 
$m_{i^*}=\tilde{m}_{i^*}-[S_{i^*}-s_{0}] -\PRF{i^*} \mod k$ and win the game, learning the reading of $SM_{i^*}$. 

In collusion set (ii), we augment the adversarial set by adding  one smart meter namely $SM_{k}$. The adversary deliberately constructs the sending list $L$ in which $SM_k$ gets to be adjacent to $SM_{i^*}$ i.e., $L=\{i^*,k,\ldots\}$.  Moreover, the adversary outputs the network failure model $G'=(V,E')$ in such a way that there is an edge between $i^*$ and $k$ i.e., $(i^*,k)\in E'$. Since $SM_{k}$ gets $S_{i^*}=s_0+s_{i^*}$ and $\DC$ gets $\tilde{m_{i^*}}=m_{i^*}+s_{i^*}+\PRF{i^*}$ from $\SM_{i^*}$, together they recover 
$m_{i^*}=\tilde{m}_{i^*}-[S_{i^*}-s_{0}] -\PRF{i^*} \mod k$ and win the game, learning the reading of $SM_{i^*}$. \qed
\end{taggedblock}

\subsection{Privacy Analysis of the Homomorphic  Aggregation Protocol}
The following theorem \ref{thm_homo} states the privacy guarantee of the protocol based on homomorphic encryption: essentially, the theorem states that the two maximal collusion sets are the ones described in Figure \ref{fig:privVisu2}.  

\begin{theorem}[Privacy of the aggregation protocol using homomorphic encryption] \label{thm_homo}
Consider the protocol \papertitle of Algorithm \ref{fig: algoGeneralV02} with privacy-preserving computations done using homomorphic encryption as described in Table \ref{tab:computations}. The following holds where privacy is as defined in the $\Unlink$ game:
\begin{itemize}
    \item[(i)] If the underlying encryption scheme is Indistinguishable under Chosen Plaintext Attack i.e., IND-CPA \cite{katz2020introduction} secure, the protocol provides privacy against an adversary consisting of $N-2$ colluding smart meters. 
    \item[(ii)] The protocol provides information-theoretic privacy against the adversary controlling $\DC$. 
 \item[(iii)] Both collusion sets above are maximal.
\end{itemize}
\end{theorem}

\begin{figure}[htb]
    \centering
    \begin{minipage}{0.3\textwidth}
    
      \begin{collusiontable}[baseline=(current bounding box.base)]
        \entityname{$\SM\ {\scriptstyle 1..N}$}
        \entityname{$\DC$}
        \nomoreentitynames

        \collusion[0.95]{informationtheoretic}
        \collusioncomment[white]{left}{\tiny $N-2$}
        \collusion{insecure}
        \nextcollusionset

        \collusion[0.95]{hard}
        \collusioncomment[white]{left}{\tiny $N-2$}
        \collusion{informationtheoretic}
        \nextcollusionset
      \end{collusiontable}
      
    \end{minipage}
          \hfill
  \begin{minipage}{0.12\textwidth}
    \centering
    \printcolorkey[baseline=(current bounding box.base)]
  \end{minipage}
        
    \caption{Privacy guarantee when homomorphic encryption is employed. Breaking privacy is computationally hard for set 1 and information-theoretic for set 2.}
    \label{fig:privVisu2}
    \vspace*{-3ex}
\end{figure}
\subsubsection{Proof of Theorem \ref{thm_homo}.(i)}
If there exists a PPT adversary $\A$ which can win $\Unlink$, then we can construct a PPT adversary $B$ who can win the IND-CPA game of the encryption scheme. 

\begin{taggedblock}{long}
\textbf{Proof Overview: }
\end{taggedblock}
In the reduction, $B$ acts as the challenger for the unlinkability game and as the adversary for the CPA game. $B$ gets the encryption public key $pk$ from the CPA challenger and hands it to the adversary. $\A$ outputs $m_0, m_1$, list $L$ and the network failure model $G'$ (note that $\A$ does not output $MList$ since the non-challenged smart meters are under its own control). Given $L$ and $G'$, $B$ checks that the challenged smart meters are part of $\Lact$ and aborts if otherwise. 
$B$ leverages the success power of $\A$  to win the CPA game by delegating the measurement assignment of challenged smart meters to the CPA challenger as explained next. $B$ crafts two message vectors $\overrightarrow{v_0}=(m_0,m_1)$ and $\overrightarrow{v_1}=(m_1,m_0)$ where each signifies an ordered assignment of $m_0$ and $m_1$ to $SM_{i^{*}}$ and $SM_{j^{*}}$, respectively. $B$ outputs $\overrightarrow{v_0}$ and $\overrightarrow{v_1}$ as part of the CPA challenge phase and obtains a cipher vector $\overrightarrow{C_b}=(c_b,c_{\hat{b}})$ where bit $b$ is selected randomly by the CPA challenger. $B$ uses $c_b$ and $c_{\hat{b}}$ as the measurements of $SM_{i^*}$ and $SM_{j^*}$, respectively. The view of $\A$ in the game is one of the followings:
1)  $SM_{i^*}$ and $SM_{j^*}$ are each other's immediate neighbors in $\Lact$, in which case $\A$ as the non-challenged smart meters, obtain the aggregate of $SM_{i^*}$ and $SM_{j^*}$ encrypted measurements, i.e. $Enc_{pk}(m_b+m_{\hat{b}})$. 
2) Otherwise, $\A$, playing the immediate neighbors of challenged smart meters in $\Lact$, obtains the individual encrypted measurements of $SM_{i^*}$ and $SM_{j^*}$, i.e. $c_b=Enc_{pk}(m_b)$ $c_{\hat{b}}=Enc_{pk}(m_{\hat{b}})$. The former case does not carry any information about $b$ (since it is independent of $b$), whereas the latter case might lead to some leakage about $b$ if the deployed encryption scheme is not CPA-secure, i.e. if $\A$ can distinguish between the messages encapsulated in $c_{b}$ and $c_{\hat{b}}$ (i.e., to distinguish between  $\overrightarrow{v_0}$ and $\overrightarrow{v_1}$). $\A$ outputs its guess as $b'$, which $B$ sends directly to the CPA challenger. If $\A$ has non-negligible advantage to win the unlinkability game so does $B$ in the CPA-game. 
\begin{taggedblock}{long}

\textbf{Formal proof}
\begin{enumerate}
    \item In the reduction $B$ plays the role of the challenger of the unlinkability game which controls $DC$, $SM_{i^*}$ and $SM_{j^*}$ whereas $\A$ plays on behalf of $N-1$ colluding smart meters. The initialization part of the aggregation protocol is run. $B$ is given the encryption key $pk$ from the outside CPA game.  The challenger gives the network graph $G=(V, E)$ as well as the $PRF$ keys $k_i$ of the non-challenged smart meters to $\A$.
    \item $\A$ outputs a pair of measurements $m_0, m_1$  within the measurement domain to be randomly assigned to two smart meters $SM_{i^*}$ and $SM_{j^*}$.
    It also sends out the network failure model $G'=(V', E')$ as well as an ordered list $L=(DC, SM_1, SM_2,\cdots, SM_N, DC)$ that indicates the arrangements of smart meters in the aggregation protocol. Note that $\A$ does not output $MList$ since the non-challenged smart meters are under its control.
    \item $B$ checks that $L$ starts and ends with $DC$, and contains all the smart meters exactly once. $B$ also checks that the challenged smart meters $SM_{i^*}$ and $SM_{j^*}$ can contribute to the aggregation in the network failure model $G'=(V', E')$ provided by $\A$, and that it is a valid failure model ($V' = V \wedge E' \subseteq E$). If any of the checks fail, $B$ aborts the game. 
    
    For the assignment of measurements to $SM_{i^*}$ and $SM_{j^*}$, $B$ proceeds as explained next. It crafts two message vectors of the following forms $\overrightarrow{v_0}=(m_0,m_1)$ and $\overrightarrow{v_1}=(m_1,m_0)$. $\overrightarrow{v_0}$  and $\overrightarrow{v_1}$  signify two different possible assignments of $m_0$ and $m_1$ to the challenged smart meters, i.e. the first and the second items in each vector represent the respective measurements of $SM_{i^{*}}$ and $SM_{j^{*}}$. $B$ submits $\overrightarrow{v_0}$ and $\overrightarrow{v_1}$ to the CPA challenger as part of the CPA challenge phase. In return, the CPA challenger picks a random bit $b$, encrypts one of the input vectors accordingly and submits the result to $B$. Let $\overrightarrow{C_b}=(c_b,c_{\hat{b}})$ denote the resultant cipher vector. Notice that if $b=0$ then $\overrightarrow{C_b}$ is the encryption of $\overrightarrow{v_0}$, and otherwise it is the encryption of $\overrightarrow{v_1}$. $B$ uses $c_b$ and $c_{\hat{b}}$ as the measurements of $SM_{i^*}$ and $SM_{j^*}$, respectively.
    \item $B$ and  $\A$ proceed with the steps of the aggregation protocol.
    \begin{enumerate}
         \item $B$ attempts the execution of line 2 of the aggregation. For each non-challenged smart meters $SM_i$ with an active link to $DC$ (i.e., $(SM_i,DC) \in E'$), $\A$ sends $\langle t,i,\{\}\rangle$ to $B$ (which plays the role of $DC$). 
        \item $B$ (as $DC$) outputs $S$ alongside with $\Lcand$, and $\Lact$ to the first smart meter in $\Lcand$ (lines 6-9 of Algorithm \ref{fig: algoGeneralV02}).
        In case that the first smart meter is one of the challenged smart meters then $B$ does this part locally.
        \item The rest of the aggregation protocol i.e., lines 10-27 of Algorithm \ref{fig: algoGeneralV02} is run. 
        During this part, $B$ as $SM_{i^*}$ (or $SM_{j^*}$) receives $\Lcand,\Lact,S$ from $\A$, and it computes $S_{i^*}=S \cdot c_b$ (or $S_{j^*}=S\cdot c_{\hat{b}}$). $B$ forwards $S_{i^*}$ (or $S_{j^*}$) to the next reachable smart meter $SM_j\in \Lcand$. To do such, $B$ checks whether $SM_{i^*}$ has an active connection link (an edge) with $SM_j$ i.e., $(SM_{i^*},SM_j)\in E'$. The presence/absence of such edge corresponds to the receipt/lack of receipt of an Ack from $SM_j$ (lines 17 and 19 of Algorithm \ref{fig: algoGeneralV02}), respectively. For smart meters with non-working links, $B$ removes them from $\Lcand$. $B$ continues with the rest of the protocol. 
        \item $B$ as $DC$ obtains $S$ and $\Lcand$ (line 29 of Algorithm \ref{fig: algoGeneralV02}). Note that $B$ does not have the decryption key, so cannot decrypt $S$. However, this does not harm its success probability in CPA game.
    \end{enumerate}
    \item  $\A$ outputs her guess $b^{'}$ to $B$. $B$ forwards $b^{'}$  directly to the CPA challenger.
\end{enumerate}

 The view of $\A$ in the game is one of the followings:
 \begin{enumerate}
     \item  $SM_{i^*}$ and $SM_{j^*}$ are each other's immediate neighbors in $\Lact$, in which case $\A$ as the non-challenged smart meters, obtain the encrypted aggregate of $SM_{i^*}$ and $SM_{j^*}$  measurements, i.e. $Enc_{pk}(m_b+m_{\hat{b}})$. This is independent of $b$ and therefore carries no information about $b$ to help $\A$ guess it correctly.
     \item Otherwise, $\A$, playing the immediate neighbors of challenged smart meters in $\Lact$ obtains the individual encrypted measurements of $SM_{i^*}$ and $SM_{j^*}$, i.e. $c_b=Enc_{pk}(m_b)$ and $c_{\hat{b}}=Enc_{pk}(m_{\hat{b}})$. To be precise, consider two colluding smart meters $SM_l$ and $SM_r$ as the left and right links of $SM_{i^*}$. $\A$ playing as $SM_r$ obtains $S_{i^*}=S_l\cdot c_{b}$ and can derive $c_b=S_{i^*} \cdot S_l^{-1}$. A similar computation yields $c_{\hat{b}}$. That is, consider $SM_{l'}$ and $SM_{r'}$ as the left and right links of $SM_{j^*}$. $SM_{r'}$ obtains $S_{j^*}=S_{l'}\cdot c_{\hat{b}}$, and also has the collusion of  $SM_{l'}$ thus knows $S_{l'}$ and can derive $c_{\hat{b}}=S_{j^*} \cdot S_{l'}^{-1}$. \\
Let denote the advantage of $\A$ in the $\Unlink$ as $\epsilon(\lambda)$ i.e., 
\begin{align}
    Pr[\Unlink=1]=\frac{1}{2}+\epsilon(\lambda)
\end{align}
Then $B$ also has the same advantage in the CPA-game. Equation \ref{eq:proof_homomorphic_N-1} illustrates this fact. $B\xrightarrow{}i$ means $B$ outputs $i$ to the CPA challenger.

\begin{align} \label{eq:proof_homomorphic_N-1}
     Pr & [ B \text{ wins CPA game}]  \nonumber \\
    = & Pr[b=0] \cdot Pr[B \xrightarrow{}0| b=0 ] + \\ \nonumber
    & Pr[b=1] \cdot Pr[B \xrightarrow{}1| b=1]  \nonumber \\
    = & \frac{1}{2}Pr[B \xrightarrow{}0| b=0 ]+\frac{1}{2}Pr[B \xrightarrow{}1| b=1]  \nonumber  \\ 
    = & \frac{1}{2}Pr[b'=0| b=0 ]+\frac{1}{2}Pr[b'=1 | b=1]  \nonumber \\
    = & Pr[b=b'] = Pr[\Unlink=1] =\frac{1}{2}+\epsilon(\lambda)\nonumber \\
\end{align}
Hence, as long as the underlying encryption scheme is CPA-secure (\ref{eq:probINDCPA}), $\epsilon(\lambda)$ needs to be negligible. \hfill $\qed$
\end{enumerate}

\end{taggedblock}
\subsubsection{Proof of Theorem \ref{thm_homo}.(ii)}
The privacy proof against $DC$ is information-theoretic, because the view of $\A$ toward the challenged smart meters is limited to their sum of measurements: regardless of how adversary influences the network failure model $G'$ and the arrangement list $L$ (under any $G'$ and $L$, both challenged smart meters are guaranteed to be part of the aggregation due to step 3 of the game $\Unlink$). $\A$ obtains the final aggregate  $A_t = m_b + m_{\hat{b}}+ \sum_{i \in \Lact/\{SM_{i^*},SM_{j^*}\}} m_{i,t}$  and can narrow it down to $m_b + m_{\hat{b}}$ by subtracting the measurements of non-challenged smart meters. However, the value of $m_b + m_{\hat{b}}$ is independent of $b$ thus it conveys no information for breaking the $\Unlink$ game. 
\subsubsection{Proof of Theorem \ref{thm_homo}.(iii)}
\begin{taggedblock}{short}
To prove that a collusion set is maximal, we show that any super set of the maximal set can launch a successful attack and break $\Unlink$.
The collusion set in Theorem \ref{thm_homo}.(i) can only be extended by adding the $\DC$. Consider an adversary who wants to learn the measurement of $\SM_{i^*}$. The adversary arranges the network in a way that the first active smart meter in the aggregation phase is $\SM_{i^*}$ and its next is $\SM_{k}$ under adversarial control, which gets to see $S_{i^*}=E_{pk}(m_{i^*})$. The adversary also controls $\DC$, who has the decryption key $sk$ and can retrieve $m_{i^*}= D_{sk}(S_{i^*})$. He wins the game by setting $b'=0$, if $m_{i^*}=m_0$ and $b'=1$ , if $m_{i^*}=m_1$. 
In collusion set (ii), if we extend the adversarial set by adding  one smart meter, namely $SM_{k}$, then the adversary can break privacy following the preceding attack scenario. \qed
\end{taggedblock}
\begin{taggedblock}{long}
To prove that a collusion set is maximal, we show that any super set of the maximal set can launch a successful attack and break $\Unlink$.
The collusion set in Theorem \ref{thm_homo}.(i) can only be extended by adding the $\DC$. Consider an adversary who wants to learn the measurement of $\SM_{i^*}$. The adversary arranges the network in a way that the first active smart meter in the aggregation phase is
$\SM_{i^*}$ and its next is $\SM_{k}$ under adversarial control,  i.e., $L=\{i^*,k,\ldots\}$. Moreover, the adversary outputs the network failure model $G'=(V,E')$ in such a way that there is an edge between $i^*$ and $k$ i.e., $(i^*,k)\in E'$. Under this arrangement, $\SM_k$  gets to see $S_{i^*}=E_{pk}(m_{i^*})$. Adversary also controls $\DC$, who has the decryption key $sk$ and can retrieve $m_{i^*}= D_{sk}(S_{i^*})$ and break the game, learning the reading of $\SM_{i^*}$.

In collusion set (ii), if we extend the adversarial set by adding  one smart meter, namely $SM_{k}$, then the adversary can break privacy following the preceding attack scenario. Essentially, the adversary  outputs sending list $L=\{i^*,k,\ldots\}$ and makes sure in its network failure model $G'$, the smart meters $i^*$ and $k$ are connected, i.e., $(i^*,k)\in E'$. During the protocol execution, $\SM_k$ obtains  $S_{i^*}=E_{pk}(m_{i^*})$ as the next of $\SM_{i^*}$. Together with the $\DC$, who holds the decryption key $sk$, they can decipher  $S_{i^*}$ and retrieve  $m_{i^*}= D_{sk}(S_{i^*})$ and win the game, learning the reading of $\SM_{i^*}$. \hfill \qed
\end{taggedblock}

\section{Conclusion and Outlook} \label{sec:conclusion}

To best of our knowledge, this is the first smart meter aggregation algorithm that can be used with both masking or homomorphic encryption. It is also the first error-resilient aggregation protocol based on these privacy preserving mechanisms with proven privacy and guaranteed fault-tolerance that is built-in without the need for further computation rounds or additional parties.
Especially when used with masking, the computational overhead is minimal.



As future work, it would be useful to develop methods for finding a maximally long path along the edges of the connectivity graph, so that the maximum number of reachable smart meters can contribute. Moreover, more detailed failure models may be analyzed and tolerated. Lastly, malicious attackers may be considered.

\begin{taggedblock}{short}
\vspace{-1cm}
\end{taggedblock}


\begin{taggedblock}{long,medium}
\section*{Acknowledgment}
We acknowledge the support of TÜBİTAK (the
Scientific and Technological Research Council of Turkey) under project number 119E088 and the Turkish Academy of Sciences. Günther Eibl gratefully acknowledges funding by the Federal State of Salzburg under the
WISS2025 program.
\end{taggedblock}

\ifCLASSOPTIONcaptionsoff
  \newpage
\fi



\nocite{*}

%

%








\end{document}